\documentclass[aps,prb,twocolumn,superscriptaddress,longbibliography]{revtex4-2}

\usepackage{natbib}
\usepackage{pgfplots}
\usepackage[breaklinks]{hyperref}
\hypersetup{
   colorlinks=true,        % false: boxed links; true: colored links
   linkcolor=red,         % color of internal links       (red)
   citecolor=blue,         % color of links to bibliography
   filecolor=blue,         % color of file links
   urlcolor=blue           % color of external links
}
\usepackage{amssymb}
\usepackage{amsmath,amsthm}
\usepackage{amsfonts,dsfont}
\usepackage{graphicx}
\usepackage{mathtools}
\usepackage{tikz}
\usepackage[caption=false]{subfig}
\usetikzlibrary{positioning}
\usepackage{accents}
\newcommand\thickbar[1]{\accentset{\rule{.5em}{.5pt}}{#1}}
\usepackage[normalem]{ulem}
\usepackage{color}
\usepackage{xcolor}

\newcommand{\bd}[1]{\boldsymbol{#1}}

\newcommand{\gh}{\hat{\gamma}}
\newcommand{\Gh}{\hat{\Gamma}}

\newcommand{\F}{\mathcal{F}}
\newcommand{\rmd}{\mathrm{d}}

\newcommand{\Tr}{\mbox{Tr}}

\newcommand{\bra}[1]{\mbox{$\langle #1 |$}}
\newcommand{\ket}[1]{\mbox{$| #1 \rangle$}}

\begin{document}
\title{Functional Theory for Bose-Einstein Condensates
% and fundamental origin of Quantum Depletion
}
\author{Julia Liebert}
\affiliation{Department of Physics, Arnold Sommerfeld Center for Theoretical Physics,
Ludwig-Maximilians-Universit\"at M\"unchen, Theresienstrasse 37, 80333 M\" unchen, Germany}
\affiliation{Munich Center for Quantum Science and Technology (MCQST), Schellingstrasse 4, 80799 M\"unchen, Germany}

\author{Christian Schilling}
\email{c.schilling@physik.uni-muenchen.de}
\affiliation{Department of Physics, Arnold Sommerfeld Center for Theoretical Physics,
Ludwig-Maximilians-Universit\"at M\"unchen, Theresienstrasse 37, 80333 M\" unchen, Germany}
\affiliation{Munich Center for Quantum Science and Technology (MCQST), Schellingstrasse 4, 80799 M\"unchen, Germany}
\affiliation{Wolfson College, University of Oxford, Linton Rd, Oxford OX2 6UD, United Kingdom}

\date{\today}

\begin{abstract}
One-particle reduced density matrix functional theory would potentially be the ideal approach for describing Bose-Einstein condensates. It namely replaces the macroscopically complex wavefunction by the simple one-particle reduced density matrix, therefore provides direct access to the degree of condensation and still recovers quantum correlations in an exact manner.
We initiate and establish this novel theory by deriving the respective universal functional $\mathcal{F}$  for homogeneous Bose-Einstein condensates with arbitrary pair interaction. Most importantly, the successful derivation
necessitates a particle-number conserving modification of Bogoliubov theory and a solution of the common phase dilemma of functional theories.
We then illustrate this novel approach in several bosonic systems such as homogeneous Bose gases and the Bose-Hubbard model. Remarkably, the general form  of $\mathcal{F}$ reveals the existence of a universal Bose-Einstein condensation force which provides an alternative and more fundamental explanation for quantum depletion.
\end{abstract}

\maketitle

\section{Introduction}\label{sec:intro}
Bose-Einstein condensation (BEC) is one of the most fascinating quantum phenomena.
While its theoretical prediction by Einstein \cite{Einstein} based on Bose's work \cite{Bose} dates back to 1925, the realization of BEC for ultracold atoms in 1995 \cite{Anderson1995,Ketterle1995,Bradley1995} has led to a renewed interest. The development of the respective field
of ultracold gases has opened new research avenues and revealed new phenomena such as the crossover from BEC-superfluidity to BCS-superconductivity \cite{Greiner2003,Bartenstein2004,Zwierlein2004,Bourdel2004}. The general need to describe bosonic quantum systems within and also beyond the ordinary BEC regime has urged us very recently to put forward a novel physical theory for describing interacting bosonic quantum systems \cite{Benavides20}. This bosonic reduced density matrix functional theory (RDMFT) is based on a generalization of the Hohenberg-Kohn theorem, abandons the complex $N$-boson wavefunction but still recovers quantum correlations in an exact way. Since it involves the one-particle reduced density matrix (1RDM) $\gh$ as the natural variable it would be particularly well-suited for the accurate description of Bose-Einstein condensates. Indeed, according to the Penrose-Onsager criterion \cite{Penrose1956}, BEC is present whenever the largest eigenvalue  of the 1RDM, $n_{max} = \max_{\varphi}\bra{\varphi}\gh\ket{\varphi}$, is proportional to the total particle number $N$. As a matter of fact, $n_{max}$  quantifies the number of condensed bosons, without requiring any preceding information about the form of the maximally populated one-particle state $\ket{\varphi_{max}}$.

While bosonic RDMFT would potentially be the ideal theory for describing BECs (including the regime of fractional BEC as well as  quasicondensation \cite{Popov1972}), RDMFT of course does not trivialize the ground state problem. Instead, it is the fundamental challenge in RDMFT to construct reliable approximations of the universal interaction functional $\F(\gh)$, determine its leading order behavior in certain physically regimes or its exact form for simplified model systems. Results along any of those lines are typically quite rare, however, and their significance for the general development of RDMFT could hardly be overestimated. The latter is due to the fact that improved functional approximations do often build upon previous ones (see, e.g., \cite{ML,PirisUg14,PG16} and references therein). In fermionic RDMFT, the elementary Hartree-Fock functional \cite{LiebHF} can be seen as the first level of the hierarchy of functional approximations. It has directly led to the celebrated M\"uller functional \cite{Mueller84,Buijse02} which in turn inspired more elaborated functional approximations \cite{ML,PG16}. In bosonic RDMFT even the analog of the Hartree-Fock functional has not been established yet.
It is therefore the main goal of our work to initiate and establish this novel bosonic RDMFT by deriving such a first-level functional in a comprehensive way. Due to the significance of BEC, we identify systems of interacting bosons in the BEC regime as the starting point for the hierarchy of functional approximations. To be more specific, we refer here to those BECs which are well described by the Bogoliubov theory.
It is worth noticing that this regime covers a large range of systems, including in particular the experimentally realized dilute ultracold Bose gases as well as charged bosons in the high density regime. The respective first-level functional may then not only serve as a starting point for the development of further functional approximations but its concrete form will also reveal a remarkable new physical concept. The gradient will namely be found to diverge repulsively in the regime of almost complete BEC, preventing quantum systems of interacting bosons from ever reaching complete condensation. This \emph{BEC force} will thus provide an alternative explanation for quantum depletion which is most \emph{fundamental}: It emerges from the geometry of density matrices and the properties of the partial trace, independently of the pair-interaction between the bosons and other system-specific features.

The paper is structured as follows. In Sec.~\ref{sec:foundation} we discuss the foundation of bosonic RDMFT, recall conventional Bogoliubov theory and explain why the latter is incompatible with RDMFT from a conceptual point of view. Then, in Sec.~\ref{sec:derivation} we present a particle-number conserving modification of Bogoliubov's theory which eventually allows us to derive the universal functional within the BEC  regime. We then illustrate in Sec.~\ref{sec:appl} how bosonic RDMFT is applied and present functionals for a number of different systems. In Sec.~\ref{sec:BECforce} we establish and illustrate the novel concept of a BEC force. In the Summary and Conclusion, we provide also a general idea for constructing higher order functional approximations based on a perturbational theoretical generalization of Bogoliubov theory.

\section{Notation and concepts}\label{sec:foundation}
We outline in this section foundational aspects of RDMFT and its application to homogeneous bosonic systems. We also introduce the most relevant concepts which will be used in subsequent sections and briefly recap conventional Bogoliubov theory.
Since our work involves rather different concepts some of which are not broadly known yet this section shall be rather comprehensive to make our
paper self-contained.

\subsection{RDMFT\label{sec:RDMFT}}
The one-particle reduced density matrix (1RDM) $\gh$ of an $N$-fermion/boson quantum state $\hat{\Gamma}$ is obtained by tracing out all except one particle
\begin{equation}\label{Gtog}
\gh \equiv N \mbox{Tr}_{N-1}[\hat{\Gamma}]\,.
\end{equation}
Equivalently, it can be characterized as the mathematically most primitive object which still determines the expectation values of all one-particle observables $\hat{h}$,
\begin{equation}
\langle \hat{h}\rangle_{\hat{\Gamma}} = \mbox{Tr}_1[\hat{h}\gh]\,.
\end{equation}
By exploiting the latter, Gilbert in 1975 \cite{Gilbert} has proven for  quantum systems of identical fermions/bosons with Hamiltonians
\begin{equation}\label{Hfamily}
\hat{H}(\hat{h}) \equiv \hat{h} + \hat{W}
\end{equation}
the existence of a universal interaction functional $\mathcal{F}_{\hat{W}}(\hat{\gamma})$ of the 1RDM: The energy and 1RDM of the ground state of $\hat{H}(\hat{h})$ for any one-particle Hamiltonian $\hat{h}$ can be determined by minimizing the total energy functional
\begin{equation}\label{energyfunc}
\mathcal{E}(\hat{\gamma}) = \Tr_1[\hat{h}\hat{\gamma}] + \mathcal{F}_{\hat{W}}(\hat{\gamma})\,.
\end{equation}
In particular, there is even a one-to-one correspondence between ground state 1RDMs and (non-degenerate) ground states \cite{Gilbert}.
In that sense, this reduced density matrix functional theory (RDMFT) is an exact ground state theory which recovers correlations in an exact manner.
The significance of RDMFT is based on the fact that the interaction functional  $\mathcal{F}_{\hat{W}}(\hat{\gamma})$ does not depend on the choice of the one-particle Hamiltonian $\hat{h}$ but only on the interaction $\hat{W}$. Since the latter is typically fixed in each scientific field (we therefore drop the index $\hat{W}$ in the following), RDMFT is a particularly economic approach for addressing the ground state problem. Indeed, any effort to approximate $\mathcal{F}(\hat{\gamma})$ contributes to the solution of the ground state problem of $\hat{H}(\hat{h})$ \emph{for all} $\hat{h}$ \emph{simultaneously}. This shall be contrasted with wavefunction-based methods whose application to $\hat{H}(\hat{h})$ does in general not provide any simplifying information towards solving other systems $\hat{H}(\hat{h}')$.

In Gilbert's RDMFT the domain of the universal functional comprises exactly those 1RDMs $\gh$ which follow from ground states, i.e., there exists an $\hat{h}$ with $\hat{H}(\hat{h}) \mapsto \ket{\Psi} \mapsto \gh$. To circumvent the problem of describing this complicated set
the constrained search formalism has been established which is based on the following consideration \cite{LE79,V80,L83}
\begin{eqnarray}\label{Levy}
E(\hat{h}) &=& \min_{\hat{\Gamma}}\,\Tr_N\left[(\hat{h} + \hat{W})\hat{\Gamma}\right] \\
&=& \min_{\gh} \min_{\hat{\Gamma}\mapsto \gh}\,\Tr_N\left[(\hat{h} + \hat{W})\hat{\Gamma}\right] \nonumber \\
&=& \min_{\hat{\gamma}}\,\Big[\Tr_1[\hat{h}\hat{\gamma}] + \underbracket{\min_{\hat{\Gamma}\mapsto\hat{\gamma}}\Tr_N[\hat{W}\hat{\Gamma}]}_{\equiv \mathcal{F}(\gh)}\Big]\nonumber \,.
\end{eqnarray}
In the last two lines, the expression $\hat{\Gamma}\mapsto \gh$ means to minimize only with respect to those $N$-fermion density operators $\Gh$ which map according to \eqref{Gtog} to the 1RDM $\gh$. This variational minimization in \eqref{Levy} can refer to either pure or ensemble $N$-particle quantum states $\hat{\Gamma}$.
Depending on that choice, the constrained search formalism leads to the pure/ensemble 1RDM-functional $\mathcal{F}$ with a domain given by all pure/ensemble $N$-representable 1RDMs. In the context of fermionic quantum systems, the complexity of the pure one-body $N$-representability conditions (generalized Pauli constraints) will thus hamper the calculation of either the functional's domain or the functional itself \cite{Schilling2018}.

\subsection{Bosonic RDMFT for homogeneous systems\label{sec:bRDMFT}}
While RDMFT has been developed and applied so far only for fermionic quantum systems, various concepts are applicable to bosonic quantum systems as well, yet with one crucial simplification. Since in case of bosons any 1RDM, $\gh \equiv \sum_\alpha\lambda_\alpha\ket{\alpha}\!\bra{\alpha}$, is pure $N$-representable (e.g.~to $\ket{\Phi}= 1/\sqrt{N}\sum_\alpha\sqrt{\lambda_\alpha}\ket{\alpha, \ldots, \alpha}$) the respective bosonic RDMFT is not hampered by one-body $N$-representability constraints. Another crucial reason for us for putting forward bosonic RDMFT \cite{Benavides20} was that the 1RDM is the crucial entity for the description and quantification of BEC \cite{Penrose1956} which is one of the most fascinating quantum phenomena. In that respect, it is worth noticing that the widely used density functional theory \cite{HK,PY95,GD95} fails to provide a direct description of BEC since its natural variable is the too rudimentary particle density.

Since our work is concerned with \emph{homogeneous} BECs, we consider only one-particle Hamiltonians which are diagonal in the momentum representation, i.e., there is only a kinetic energy operators $\hat{t}$ contributing to $\hat{h}$ but no external potential, $\hat{h}\equiv \hat{t}$. Implementing this within the constrained search formalism \eqref{Levy} identifies the momentum occupation numbers $\bd{n}\equiv (n_{\bd{p}})$ as the natural variables and the pure functional follows as
\begin{equation}\label{LevyP}
\F(\bd{n}) \equiv \min_{\ket{\Phi}\mapsto \bd{n}}\bra{\Phi}\hat{W}\ket{\Phi}\,.
\end{equation}
While we are focussing in the following on the pure functional, it is worth recalling that the corresponding ensemble functional would follow as the lower convex envelop of the pure functional $\F$ \cite{Schilling2018}. Also their two domains $\triangle$ coincide.
To describe $\triangle$, let us first use the normalization constraint to get rid of the entry $n_\textbf{0}= N-\sum_{\bd{p}\neq \textbf{0}} n_{\bd{p}}$.
Then the functional's domain follows as
\begin{equation}\label{simplex}
\triangle = \Big\{\bd{n} \equiv (n_{\bd{p}})_{\bd{p}\neq \textbf{0}} \Big| n_{\bd{p}}\geq 0, \sum_{\bd{p}\neq \textbf{0}} n_{\bd{p}}\leq N \Big\}\,.
\end{equation}
In case of finite lattice models there are finitely many momenta $\bd{p}$ (forming a discrete Brillouin zone), while in case of continuous systems or infinite lattices, $\bd{n}$ will have infinitely many entries.
It will be instructive to also understand the functional's domain $\triangle$ from a geometric point of view.
Apparently, $\triangle$ is a convex set which after all takes the form of a simplex with vertices $\textbf{0}$ and $\bd{v}_{\bd{p}} = N \bd{e}_{\bd{p}}$, where $\bd{e}_{\bd{p}}$ has only one non-vanishing entry $1$ at position $\bd{p}$. In the following, we are mainly interested in the regime of BEC which is characterized by an occupation number $n_{\textbf{0}}$ close to $N$. This corresponds in the simplex $\triangle$  to the neighborhood of the vertex $\textbf{0}$, which can equivalently be characterized by the simultaneous saturation of the constraints $n_{\bd{p}}\geq 0$ for all $\bd{p}\neq \textbf{0}$.

Last but not least, we would like to stress that the parity-symmetry of common physical spaces implies the additional symmetry $n_{\bd{p}}= n_{-\bd{p}}$ for all momenta $\bd{p}$. This does not really change the geometric form of the functional's domain but just allows us to skip in the definition \eqref{simplex} for every pair $(\bd{p},-\bd{p})$ of momenta one of the two occupation numbers $n_{\pm \bd{p}}$.
In the context of RDMFT, respecting this common symmetry would mean to restrict the kinetic energy operators $\hat{t} \equiv \sum_{\bd{p}}\varepsilon_{\bd{p}}\hat{n}_{\bd{p}}$ to those with $\varepsilon_{\bd{p}}=\varepsilon_{-\bd{p}}$.

\subsection{Recap of conventional Bogoliubov theory \label{sec:Bog}}
In this section we recap the most important aspects of Bogoliubov's \cite{Bogoliubov47} well-known and experimentally confirmed \cite{Lopes17} theory to describe BEC in homogenous bosonic quantum systems and the effect of depletion of the condensate as a result of the  interaction.

The Hamiltonian describing a homogenous system of $N$ interacting spinless bosons in first quantization ($\hbar\equiv 1$) is given by
\begin{equation}\label{H1st}
\hat{H} = -\sum_{i=1}^N\frac{1}{2m}\Delta_i + \sum_{1\leq i<j\leq N}W(\bd{x}_i-\bd{x}_j)\,.
\end{equation}
Its second quantized form in momentum representation for particles in a large box of volume $V=L^3$ and size $L$ with periodic boundary conditions then reads
\begin{equation}\label{H2nd}
\hat{H} = \sum_{\bd{p}}\varepsilon_{\bd{p}}\hat{a}_{\bd{p}}^\dagger\hat{a}_{\bd{p}} + \frac{1}{2V}\sum_{\bd{p}, \bd{q}, \bd{k}} W_{\bd{p}} \hat{a}_{\bd{p}+\bd{q}}^\dagger\hat{a}_{\bd{k}-\bd{p}}^\dagger\hat{a}_{\bd{k}}\hat{a}_{\bd{q}}\,,
\end{equation}
where $W_{\bd{p}}$ is the Fourier transform of $W(\cdot)$. In case of an isotropic pair interactions, $W$ in Eq.~\eqref{H1st} would depend only on the modulus of the distance between the particles $i$ and $j$ which in turn would imply $W_{\bd{p}}\equiv W_{|\bd{p}|}$.

The most crucial feature of the Hamiltonian $\hat{H}$ and the pair interaction $\hat{W}$ is that they are conserving the particle number as well as the total momentum. Assuming a BEC at $T=0$, the standard approach to determine the ground state energy (and the low lying excited states) of the Hamiltonian \eqref{H2nd} is the Bogoliubov approximation \cite{Bogoliubov47}. It is based on the fact that in the regime of BEC the zero-momentum mode is macroscopically occupied and interactions between non-condensed bosons can be neglected  due to the conservation of momentum: Since application of a creation/annihilation operator $a_{\mathbf{0}}^{(\dagger)}$ to the BEC ground state leads to macroscopically large prefactors of the order $\sqrt{N}$, terms in the expansion \eqref{H2nd} of $\hat{W}$ involving less than two $\bd{0}$-indices are dropped. The resulting quartic interaction is further simplified by replacing the condensate operators $\hat{a}_{\mathbf{0}}, \hat{a}_{\mathbf{0}}^\dagger \to \sqrt{n_{\mathbf{0}}} \approx\sqrt{N}$ by a c-number. This eventually leads to the quadratic Bogoliubov Hamiltonian which involves (besides the kinetic energy $\hat{t}$ and some trivial contributions) for each pair $(\bd{p},-\bd{p})$ an anomalous term of the form $\hat{a}_{\bd{p}}^\dagger\hat{a}_{-\bd{p}}^\dagger + \hat{a}_{\bd{p}}\hat{a}_{-\bd{p}}$.
The Bogoliubov Hamiltonian can then easily be diagonalized by a Bogoliubov transformation $\hat{U}_B = \mathrm{exp}\left\{\frac{1}{2}\sum_{\bd{p}\neq 0} \theta_{\bd{p}}\left(\hat{a}_{\bd{p}}^\dagger\hat{a}_{-\bd{p}}^\dagger - \mathrm{h.c.}\right)\right\}$. The respective ground state follows as $\ket{\Psi} = \hat{U}_B\ket{N}$, where $\ket{N} \equiv (N!)^{-1/2}(\hat{a}_0^\dagger)^N\ket{0}$ is the ground state of the non-interacting system and $\ket{0}$ the vacuum state. The phases $\theta_{\bd{p}}$ are chosen such that the anomalous terms in the Hamiltonian, containing either two quasiparticle annihilation ($\hat{b}_{\bd{p}}\equiv \hat{U}_B^\dagger \hat{a}_{\bd{p}}\hat{U}_B$) or creation operators ($\hat{b}_{\bd{p}}^\dagger$) vanish to eventually obtain a  diagonal quadratic form in $\hat{b}_{\bd{p}}$ (see also textbook \cite{pita} for more details).
Bogoliubov's approach can also be interpreted as the variational minimization of the Bogoliubov Hamiltonian over all trial states of the form $\hat{U}_B\ket{N}$.

\subsection{Incompatibility of RDMFT and conventional Bogoliubov theory\label{sec:obstacles}}
As explained in the previous section, Bogoliubov's approximation results in a Hamiltonian which is not particle-number conserving anymore.
At the same time, RDMFT defines a universal functional $\F(\bd{n})$ (or more generally $\F(\gh)$) by minimizing the interaction Hamiltonian according to \eqref{LevyP} with respect to quantum states with a \emph{fixed} total particle number $N$ and fixed momentum occupation numbers $\bd{n}$. Replacing in Eq.~\eqref{LevyP} $\hat{W}$ by Bogoliubov's approximated Hamiltonian would therefore erroneously ignore the important anomalous terms  $\hat{a}_{\bd{p}}^\dagger\hat{a}_{-\bd{p}}^\dagger + \hat{a}_{\bd{p}}\hat{a}_{-\bd{p}}$. At first sight, this incompatibility of Bogoliubov's conventional approximation and RDMFT seems to be paradoxical. Yet, it is worth recalling that the merits of the unitary Bogoliubov transformation lie in the simple calculation of the (low-lying) energy spectrum while its violation of particle-number conservation can lead to conceptual difficulties beyond RDMFT as well. At the same time, since RDMFT has the distinctive goal to (partly) solve the ground state problem for $\hat{H}(\hat{h})$  for all $\hat{h}$ simultaneously, it requires apparently a mathematically more rigid and well-defined framework than the one  provided by conventional Bogoliubov theory.

Before we discuss in the following section such a well-defined mathematical framework for realizing Bogoliubov's ideas within RDMFT, we briefly comment on an alternative natural idea for circumventing the outlined difficulties. Instead of applying the constrained search formalism to a fixed particle number sector, one could also extend \eqref{LevyP} to the entire Fock space. This would result in a Fock space RDMFT and the anomalous terms would contribute to the functional. Yet, there would be a crucial drawback. The respective functional would namely allow one for any Hamiltonian \eqref{Hfamily} to only calculate the \emph{overall} ground state on the Fock space. For instance, for specific kinetic energy operators or pair interactions, this overall minimum may lie in the sector of zero or infinitely many bosons. Also adding a chemical potential term $\mu \hat{N}$ for steering the particle number to a preferred one would only work in case the Fock space functional was convex in the total particle number.

\section{Derivation of the universal functional\label{sec:derivation}}

\subsection{Particle-number conserving Bogoliubov theory \label{sec:pair}}
As discussed in the context of Eq.~\eqref{Levy} and motivated in Sec.~\ref{sec:obstacles}, the derivation of the universal functional for BECs requires a particle-number conserving variant of conventional Bogoliubov theory. Exactly such a modification has been provided by
Girardeau \cite{Girardeau59} (see also \cite{Gardiner97,Girardeau98,Seiringer11,Seiringer14}) in the context of pair theory. In the following, we will outline and then apply this theory which in particular improves upon Bogoliubov theory by including more terms of the Hamiltonian.
The idea behind pair theory is that in the regime of BEC, excitations of pairs $(\bd{p}, -\bd{p})$ from the condensate dominate and thus the interacting ground state is well approximated by a state with a corresponding pairing structure \cite{Girardeau59, Girardeau98}. Restricting the original Hamiltonian $\hat{H}$ to the space of such pair excitation states and assuming that the zero-momentum state is macroscopically occupied means to effectively deal with a modified interaction $\hat{W}_P$ of pair excitation type \cite{Girardeau59},
\begin{widetext}
\begin{eqnarray}\label{WP}
%\hat{W}|_{\mathcal{H}_P}  \equiv
\hat{W}_P &\equiv& \frac{N(N-1)W_{\mathbf{0}}}{2V} + \frac{1}{2V}\sum_{\bd{p}\neq 0}W_{\bd{p}}\left[2\hat{n}_{\mathbf{0}}\hat{n}_{\bd{p}} +\hat{a}_{\bd{p}}^\dagger\hat{a}_{-\bd{p}}^\dagger \hat{a}_{\mathbf{0}}^2 + \big(\hat{a}_{\mathbf{0}}^\dagger\big)^2\hat{a}_{\bd{p}}\hat{a}_{-\bd{p}} \right] \\
&&+ \frac{1}{2V}\sum_{\substack{\bd{p}, \bd{p^\prime}\neq 0 \\ \bd{p}\neq \bd{p}^\prime}} W_{\bd{p}}\hat{a}_{\bd{p}^\prime}^\dagger\hat{a}_{-\bd{p}^\prime}^\dagger\hat{a}_{\bd{p}^\prime-\bd{p}}\hat{a}_{\bd{p}-\bd{p}^\prime} + \frac{1}{2V}\sum_{\substack{\bd{p},\bd{p^\prime}\neq 0 \\ \bd{p}\neq \bd{p}^\prime, \bd{p}\neq 2\bd{p}^\prime}} W_{\bd{p}}\hat{n}_{\bd{p}^\prime-\bd{p}}\hat{n}_{\bd{p}^\prime}\nonumber\,.
\end{eqnarray}
\end{widetext}
The terms in the first line of Eq.~\eqref{WP} give rise to the Bogoliubov Hamiltonian (after the replacement $\hat{a}_{\mathbf{0}}, \hat{a}_{\mathbf{0}}^\dagger \to\sqrt{N}$) while those in the second line improve upon Bogoliubov theory.
For the sake of simplicity, we omit in the following derivations the constant term $\frac{N(N-1)W_{\mathbf{0}}}{2V}$.
To determine a variational ground state energy of $\hat{H}(\hat{h})$, Girardeau's idea was then to employ a particle-number conserving analogue of Bogoliubov trial state $\hat{U}_B\ket{N}$. For this, one first introduces the operators
\begin{equation}\label{beta0}
\hat{\beta}_{\mathbf{0}} \equiv \left(\hat{n}_{\mathbf{0}}+1\right)^{-1/2}\hat{a}_{\mathbf{0}}\,,\quad
\hat{\beta}_{\mathbf{0}}^\dagger \equiv \hat{a}_{\mathbf{0}}^\dagger  \left(\hat{n}_{\mathbf{0}}+1\right)^{-1/2}
\end{equation}
which annihilate/create a boson in the condensate, yet without changing the normalization of the respective quantum state.
Girardeau's $N$-boson trial states
\begin{equation}\label{Psi_int}
|\Psi\rangle \equiv \hat{U}_G|N\rangle
\end{equation}
of pair excitation form are defined by the following operators
\begin{equation}\label{U}
\hat{U}_G = \mathrm{exp}\left\{\frac{1}{2}\sum_{\bd{p}\neq 0}\theta_{\bd{p}}\left[\left(\hat{\beta}_{\mathbf{0}}^\dagger\right)^2\hat{a}_{\bd{p}}\hat{a}_{-\bd{p}} - \hat{\beta}_{\mathbf{0}}^2\hat{a}_{\bd{p}}^\dagger\hat{a}_{-\bd{p}}^\dagger\right]\right\}
\end{equation}
with $\theta_{\bd{p}}\in \mathbb{R}$ and $\theta_{\bd{p}} = \theta_{-\bd{p}}$.
The operators $\hat{U}_G$ are particle-number conserving as desired, $[\hat{U}_G, \hat{N}]=0$, which is due to the additional operators $\hat{\beta}_{\mathbf{0}}$ and $\hat{\beta}_{\mathbf{0}}^\dagger$.
Since its exponent is antihermitian, $\hat{U}_G$ is still unitary (as $\hat{U}_B$). The price one has to pay for the more complicated exponent, however, is that no compact exact expression can be found for the quasiparticle operators $\hat{U}_G^\dagger\hat{a}_{\bd{p}}\hat{U}_G$ anymore. Instead the result known from Bogoliubov theory holds only approximately,
\begin{equation}\label{qp}
\hat{U}_G^\dagger\hat{a}_{\bd{p}}\hat{U}_G \approx \frac{1}{\sqrt{1-\phi_{\bd{p}}^2}}\left(\hat{a}_{\bd{p}} - \phi_{\bd{p}}\beta_{\mathbf{0}}^2\hat{a}_{-\bd{p}}^\dagger\right) \equiv \hat{\xi}_{\bd{p}}\,,
\end{equation}
where $\phi_{\bd{p}} \equiv \tanh (\theta_{\bd{p}})$. A careful mathematical estimate of the difference between left and right side of \eqref{qp} has been provided in \cite{Seiringer11} (yet involving a slightly different but conceptually similar definition of $\hat{U}_G$).
It effectively allows us to treat \eqref{qp} and the implied Eq.~\eqref{np} as exact relations for our further derivation.
The particle number expectation value of the momentum mode $\bd{p}\neq 0$ in the interacting ground state $\ket{\Psi}$ \eqref{Psi_int} then follows as
\begin{equation}\label{np}
n_{\bd{p}} \equiv \langle\Psi|\hat{a}_{\bd{p}}^\dagger\hat{a}_{\bd{p}}|\Psi\rangle \approx \langle N|\hat{\xi}_{\bd{p}}^\dagger\hat{\xi}_{\bd{p}}|N\rangle = \frac{\phi_{\bd{p}}^2}{1-\phi_{\bd{p}}^2}\,.
\end{equation}

\subsection{Calculation of the Functional\label{sec:derivfunc}}
Relation \eqref{np} is the crucial ingredient for our derivation of the universal functional $\F$ in the regime of BEC. This connection between the family of variational trial states of fixed particle number and the momentum occupation numbers $\bd{n}$ will drastically simplify the constrained search \eqref{LevyP} and will allow us eventually to determine the explicit form of $\F$.
For this, we observe that relation \eqref{np} can be inverted up to binary degrees of freedom $\sigma_{\bd{p}}= \sigma_{-\bd{p}}=\pm 1$,
\begin{equation}
\phi_{\bd{p}} = \sigma_{\bd{p}} \sqrt{\frac{n_{\bd{p}}}{1+n_{\bd{p}}}}\,.
\end{equation}
This sign ambiguity is conceptually very similar to the so-called phase dilemma in fermionic RDMFT \cite{PernalPhase04}. The latter resembles the fact that general phase changes of the natural orbitals (eigenstates of the 1RDM) affect $\bra{\Psi}\hat{W}\ket{\Psi}$ in \eqref{Levy} via the $N$-fermion wavefunction $\ket{\Psi}$ while keeping the 1RDM invariant. In contrast to fermionic RDMFT, however, the minimizing signs $\{\sigma_{\bd{p}}\}$ can be found in our case of bosons in the BEC regime.

We combine now various concepts and ideas to determine the universal functional $\F(\bd{n})$ for BECs.
According to the constrained search formalism \eqref{LevyP} we need to minimize for any vector $\bd{n}$ the expectation value of the interaction $\hat{W}$ over all $N$-boson quantum states with momentum occupation numbers $\bd{n}$. Our focus on the regime of BECs then allows us to restrict this to Girardeau's $N$-boson trial states \eqref{Psi_int} with the additional effect that $\hat{W}$ simplifies to $\hat{W}_P$ in Eq.~\eqref{WP}, i.e., $\langle\Psi|\hat{W}|\Psi\rangle =\langle\Psi|\hat{W}_P|\Psi\rangle = \bra{N}\hat{U}_G^\dagger W_P\hat{U}_G\ket{N}$. The operator $\hat{U}_G^\dagger W_P\hat{U}_G$ should then be expressed in terms of the quasiparticle operators $\hat{\xi}_{\bd{p}}$ given by Eq.~\eqref{qp}, allowing us to eventually calculate its action on the state $\ket{N}$.
Since the trial states $\ket{\Psi}$ are almost uniquely determined by $\bd{n}$ according to \eqref{np}
we are only left with a minimization over all possible combinations of signs $\sigma_{\bd{p}}$.
Keeping only terms which do not vanish in the thermodynamic limit $N\to \infty$, $V\to \infty$ and $n=N/V=\mathrm{cst.}$ yields
then the final result for the Girardeau approximated functional
\begin{widetext}
\begin{equation}\label{Fmin}
\mathcal{F}_G(\bd{n})% =\min_{\{\sigma_{\bd{p}}=\pm 1\}} \mathcal{F}_G^{(\bd{\sigma})}(\bd{n})
= \min_{\{\sigma_{\bd{p}}=\pm 1\}}\Bigg\{ \sum_{\bd{p}\neq 0}\Big[\frac{n_{\mathbf{0}}}{V}  W_{\bd{p}} + \frac{1}{2}I_2(\bd{p}, \bd{n})\Big]n_{\bd{p}} -\sigma_{\bd{p}}\Big[\frac{n_{\mathbf{0}}}{V} W_{\bd{p}} - \frac{1}{2}I_1(\bd{p}, \bd{n},\bd{\sigma})\Big]\sqrt{n_{\bd{p}}(n_{\bd{p}}+1)}\Bigg\} \,.
\end{equation}
\end{widetext}
where
\begin{eqnarray}\label{Iterms}
I_1(\bd{p}, \bd{n},\bd{\sigma}) &\equiv& \frac{1}{V}\sum_{\substack{\bd{p^\prime}\neq 0}}W_{\bd{p}-\bd{p}^\prime}\sigma_{\bd{p}^\prime}\sqrt{n_{\bd{p}^\prime}(n_{\bd{p}^\prime}+1)}\nonumber \\
I_2(\bd{p}, \bd{n}) &\equiv&\frac{1}{V}\sum_{\substack{\bd{p^\prime}\neq 0}}W_{\bd{p}-\bd{p}^\prime}n_{\bd{p}^\prime}\,.
\end{eqnarray}
These formulas could directly be copied from \cite{Girardeau59} since there the expectation value of the full Hamiltonian $\hat{t}+\hat{W}_P$ was calculated for the same quantum state $\hat{U}_G\ket{N}$.
For general $\bd{n} \in \triangle$ one cannot overcome the common phase dilemma and in particular the minimizing sign factors $\sigma_{\bd{p}}=\pm1$ in \eqref{Fmin} depend on $\bd{n}$. This in turn leads to a partitioning of the functional's domain $\triangle$ into cells characterized by different signs $\{\sigma_{\bd{p}}\}$, similarly to the Ising cells corresponding to different spin configurations (see also Fig.~\ref{fig:S} for an illustration). As already explained in Sec.~\ref{sec:pair}, Girardeau's approach based on pair theory goes beyond Bogoliubov theory by including additional terms of the original Hamiltonian (see also second line of \eqref{WP}). Yet, since those involve fewer creation/annihilation operators $a_{\mathbf{0}}^{(\dagger)}$ and since the Girardeau approach uses at the end (almost) the same trial states \eqref{Psi_int} as Bogoliubov, we expect that the additional terms $I_1, I_2$ in \eqref{Fmin} have only a minor quantitative rather than a significant qualitative effect on the description of BECs. Whether this changes  beyond the regime of BEC is not clear since one still restricts to the common BEC trial states \eqref{Psi_int}.

In the regime of BEC the two terms in Eq.~\eqref{Fmin} involving $I_1$ and $I_2$, respectively, are significantly smaller than the term proportional to $n_{\mathbf{0}}$. Accordingly, in the regime of interest the minimization of various $\sigma_{\bd{p}}$ can be executed analytically, leading to
\begin{equation}\label{sgnmagic}
\sigma_{\bd{p}} = \mathrm{sgn}(W_{\bd{p}})\,,\quad\forall\,\,\bd{p}\neq 0\,.
\end{equation}
Also, the possible approximation $n_{\mathbf{0}}\approx N$ would be of the same order as neglecting the less significant Girardeau terms $I_1$ and $I_2$. Eventually, implementing those two last approximations leads to one of our key results, the  Bogoliubov approximated functional ($n \equiv N/V$)
\begin{equation}\label{FBog}
\mathcal{F}_B(\bd{n}) = n\sum_{\bd{p}\neq 0}W_{\bd{p}}\left[n_{\bd{p}} - \mathrm{sgn}(W_{\bd{p}})\sqrt{n_{\bd{p}}(n_{\bd{p}}+1)}\right]\,.
\end{equation}
For the sake of completeness, we would like to emphasize that a simplified version of \eqref{FBog} has already been presented in
Ref.~\cite{Benavides20}. There, by reverse engineering we calculated this functional as the Legendre-Fenchel transform of the well-known result for the ground state energy of a Bogoliubov-approximated system. As the application of \eqref{FBog} in Sec.~\ref{subsec:dilute} will reveal, the underlying replacement $W_p \mapsto W_0\geq 0$ in \cite{Benavides20} is, however, too restrictive and also rather problematic.
The distinctive form of the Bogoliubov functional $\mathcal{F}_B$ in \eqref{FBog} resembles clearly the decoupling of various momentum pairs $(\bd{p},-\bd{p})$ from each other within Bogoliubov theory. Remarkably, the Bogoliubov approximated functional $\mathcal{F}_B$ is convex, in contrast to common pure functionals in fermionic RDMFT. The pure functional $\mathcal{F}_B$ therefore coincides with the corresponding ensemble functional since the latter is given by the lower convex envelop of the former \cite{Schilling2018}.

We also would like to reiterate that due to the general significance of BECs, the functional \eqref{FBog} can be seen as the first-level approximation of the universal functional in bosonic RDMFT. In analogy to the Hartree-Fock \cite{LiebHF} and the M\"uller functional \cite{Mueller84,Buijse02} in fermionic RDMFT and the local density approximation in density functional theory \cite{PerdewJacob}, $\mathcal{F}_B$ and $\mathcal{F}_G$ will represent a promising starting point for the construction of more elaborated functional approximations. In that sense, we expect that our key results \eqref{Fmin} and \eqref{FBog} will initiate and establish eventually bosonic RDMFT. In the following, we simplify our notation by skipping the index $B,G$ of the functional, also since both functionals (almost) coincide in the relevant regime of BEC.

\section{Illustration and Applications\label{sec:appl}}
\subsection{Dilute Bose gas in 3D\label{subsec:dilute}}
We now apply the concepts of RDMFT to the homogenous dilute Bose gas, the system for which Bogoliubov's theory \cite{Bogoliubov47} was originally developed. This will also allow us to demonstrate how the well-known expression for the ground state energy of a dilute Bose gas \cite{Brueckner57} can be obtained using RDMFT. Our derivation will be based on the s-wave scattering approximation, involving the first two terms of the Born series.

Let us introduce for the following considerations the degree $D$ of quantum depletion ($N_\mathrm{BEC} \equiv n_{\mathbf{0}}$)
\begin{equation}\label{Dgen}
D\equiv1-N_\mathrm{BEC}/N = \frac{1}{N}\sum_{\bd{p}\neq 0}n_{\bd{p}}\,.
\end{equation}
From a geometric point of view, $D$ is nothing else than the $l_1$-distance of $\bd{n}$ in the simplex $\triangle$ \eqref{simplex} to the vertex $\mathbf{0}$ corresponding to complete BEC. We also recall that the ground state energy of $\hat{H}(\hat{t})= \hat{t}+\hat{W}$ follows in RDMFT by minimizing the respective
energy functional over the space of occupation number vectors $\bd{n}$,
\begin{equation}\label{E0min}
E(\hat{t}) = \min_{\bd{n}\in \triangle}\left[\bd{\varepsilon} \cdot \bd{n} + \mathcal{F}(\bd{n})\right]\,,
\end{equation}
where $\hat{t}\equiv \sum_{\bd{p}}\varepsilon_{\bd{p}} \hat{n}_{\bd{p}}$, assuming w.l.o.g.~$\varepsilon_{\mathbf{0}} =0$, and $\bd{\varepsilon} \cdot \bd{n}\equiv \sum_{\bd{p}\neq \mathbf{0}} \varepsilon_{\bd{p}}n_{\bd{p}}$. To calculate for the \emph{realistic} dilute Bose gas the ground state energy and the degree of condensation, we would need to plug in for the kinetic energy in \eqref{E0min} the specific dispersion relation of free particles, i.e., $\varepsilon_{\bd{p}}=p^2/2m$ (ignoring boundary effects).  It is worth reiterating that in principle systems with any kinetic energy  $\hat{t}$ could be considered in RDMFT. From an experimental point of view, one could indeed imagine a modified dispersion relation due to a specific background medium and in case of lattice models both the rate and range of the hopping can be actually varied (see, e.g., Refs.~\cite{Guenter2013,Schempp2015}). Because of this, we are for the moment still considering a general $\hat{t}$ and $\bd{\varepsilon}$, respectively.

Finding the minimizer $\thickbar{\bd{n}}$ of the energy functional then means to solve
\begin{equation}\label{varE0}
\varepsilon_{\bd{p}} = -\frac{\partial \mathcal{F}}{\partial n_{\bd{p}}}(\thickbar{\bd{n}})\,,\quad \forall \bd{p}\,.
\end{equation}
Using the explicit form of the Bogoliubov functional \eqref{FBog} then leads to
\begin{equation}\label{nbar}
\thickbar{n}_{\bd{p}} = \frac{1}{2}\left(\frac{\varepsilon_{\bd{p}} + nW_{\bd{p}}}{\sqrt{\varepsilon_{\bd{p}}(\varepsilon_{\bd{p}}+2nW_{\bd{p}}})} - 1\right)\,.
\end{equation}
This is nothing else than the well-known result for the momentum occupation numbers \cite{pita}.

Considering now the specific case of a realistic dilute Bose gas then allow us to determine the ground state energy explicitly. For this, we first evaluate the universal functional at the minimum $\thickbar{\bd{n}}$, leading to (see Appendix \ref{app:Fdil})
\begin{equation}\label{Fdil}
\mathcal{F}(\thickbar{\bd{n}}) = \frac{128\sqrt{\pi}}{3m}Na_0^{5/2}n^{3/2} + \frac{4\pi Na_1 n}{m}\,.
\end{equation}
It depends only on the first two terms of the Born series of the s-wave scattering length $a$ \cite{Brueckner57},
\begin{equation}\label{a0}
a_0 = \frac{mW_{\mathbf{0}}}{4\pi} \,,\quad a_1 = -\frac{1}{4\pi V}\sum_{\bd{p}\neq 0}\frac{W_{\bd{p}}^2m^2}{p^2}\,.
\end{equation}

As it is shown in Appendix \ref{app:Fdil}, adding the kinetic energy and reintroducing the omitted constant term $W_{\mathbf{0}}N(N-1)/2V\approx W_{\mathbf{0}}Nn/2$ leads to the well-known ground state energy \cite{Brueckner57}:
\begin{eqnarray}\label{E0dilute}
E &=& \frac{2\pi Nn}{m}\left(a_0+a_1 +\frac{128}{15\sqrt{\pi}}a_0(na_0^3)^{1/2}\right)\,.
%\\&=& \frac{2\pi Nn a}{m}\left(1 +\frac{128}{15\sqrt{\pi}}(n a^3)^{1/2}\right)\left[1+\mathcal{O}\big((n a^3)^{1/2}\big)\right]\,.\nonumber
\end{eqnarray}
This can be recast by using the scattering length $a$ which eventually leads (up to higher order terms) to the compact expression \cite{pita}  $E=\frac{2\pi Nn a}{m}\left(1 +\frac{128}{15\sqrt{\pi}}(n a^3)^{1/2}\right)$.

Since the underlying domain $\triangle$ is infinite dimensional in case of a continuous system in a box it is difficult to graphically illustrate the Bogoliubov functional. Yet, to visualize at least some of its most crucial features we define two paths within $\triangle$, both starting at the physical point $\thickbar{\bd{n}}$ (corresponding to $\varepsilon_{\bd{p}} = \bd{p}^2/2m$) and terminating at the vertex $\mathbf{0}$ describing complete BEC. The first one is just the straight path $s$ between those two points, parameterized by $t\in[0,1]$,
\begin{equation}\label{nt}
\bd{n}(t) = \thickbar{\bd{n}} - t\thickbar{\bd{n}} \,.
\end{equation}
The $l_1$-distance $D(t)$ of $\bd{n}(t)$ to $\mathbf{0}$ follows directly as
\begin{equation}
D(t)=\frac{\sqrt{n}}{3\pi^2}(m W_{\mathbf{0}})^{3/2}(1-t)
\end{equation}
and the functional's concrete values $\mathcal{F}[\bd{n}(t)]$ along that path can easily be calculated by exact numerical means.
As second path $\kappa$, we consider the one experimentally realized in Ref.~\citep{Lopes17} by continuously reducing the coupling strength $\kappa$ of the pair interaction from one to zero. Since the interaction Hamiltonian $\hat{W}$ in RDMFT is fixed, this path has to be realized equivalently  by increasing the strength of the kinetic energy according to $p^2/2m\kappa$.
The respective distance $D$ of $\bd{n}(\kappa)$ to $\mathbf{0}$ follows as
\begin{equation}
D(\kappa)=\frac{\sqrt{n}}{3\pi^2}(mW_{\mathbf{0}}\kappa)^{3/2}
\end{equation}
and the functional $\mathcal{F}[\bd{n}(\kappa)]$ along that path is given by
\begin{equation}\label{Fdilkappa}
\begin{split}
\mathcal{F}[\bd{n}(\kappa)] &= 4nNW_{\mathbf{0}}D(\kappa) \\\
&\quad+ \frac{3^{2/3}4\pi^{7/3}Nn^{2/3}a_1}{m^2W_{\mathbf{0}}}D^{2/3}(\kappa)\,.
\end{split}
\end{equation}
In Eq.~\eqref{Fdilkappa} one may replace $W_{\mathbf{0}}$ by $a_0$ according to \eqref{a0}.

\begin{figure}[htb]
\includegraphics[width=0.49\linewidth]{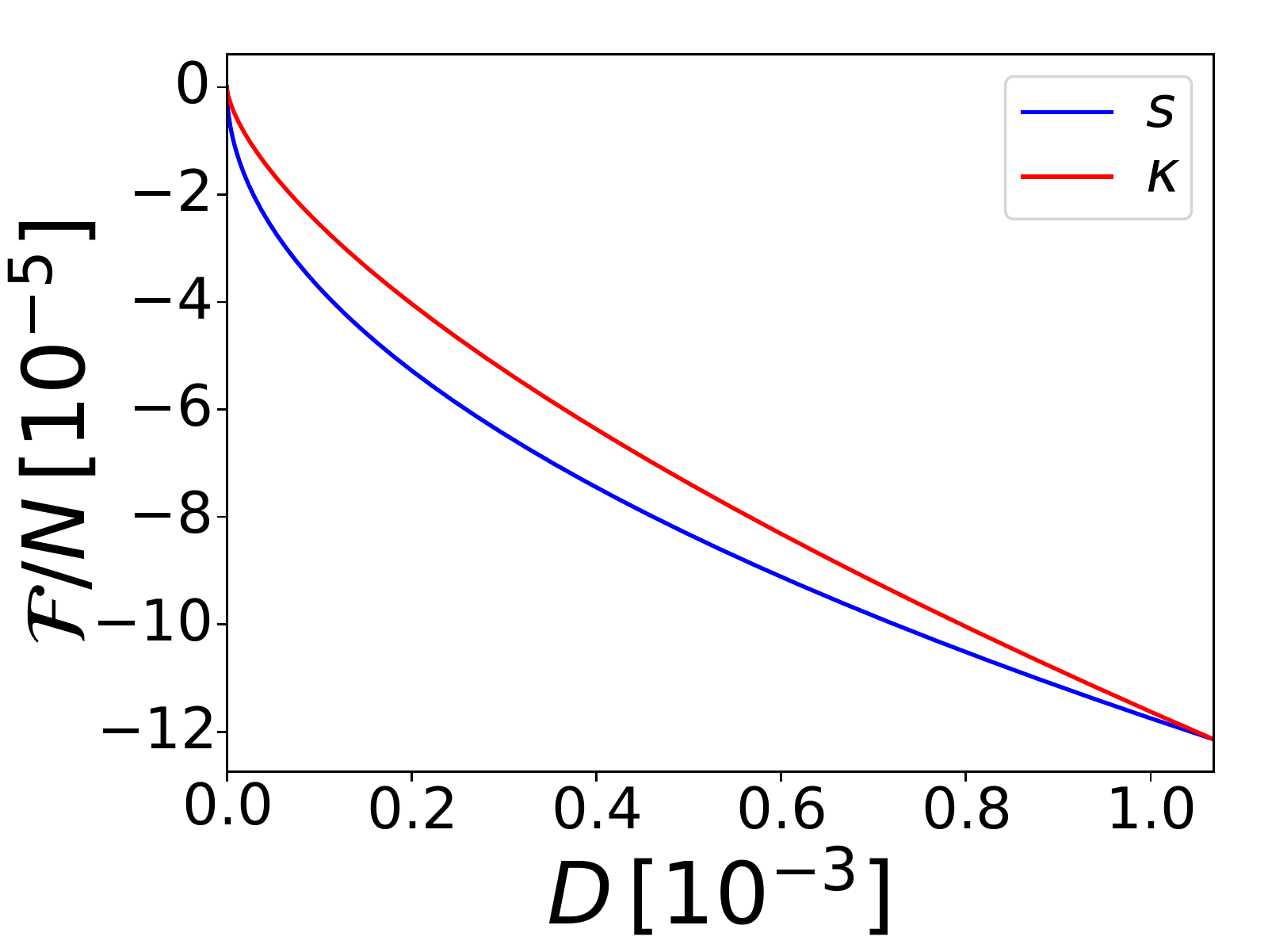}
\includegraphics[width=0.49\linewidth]{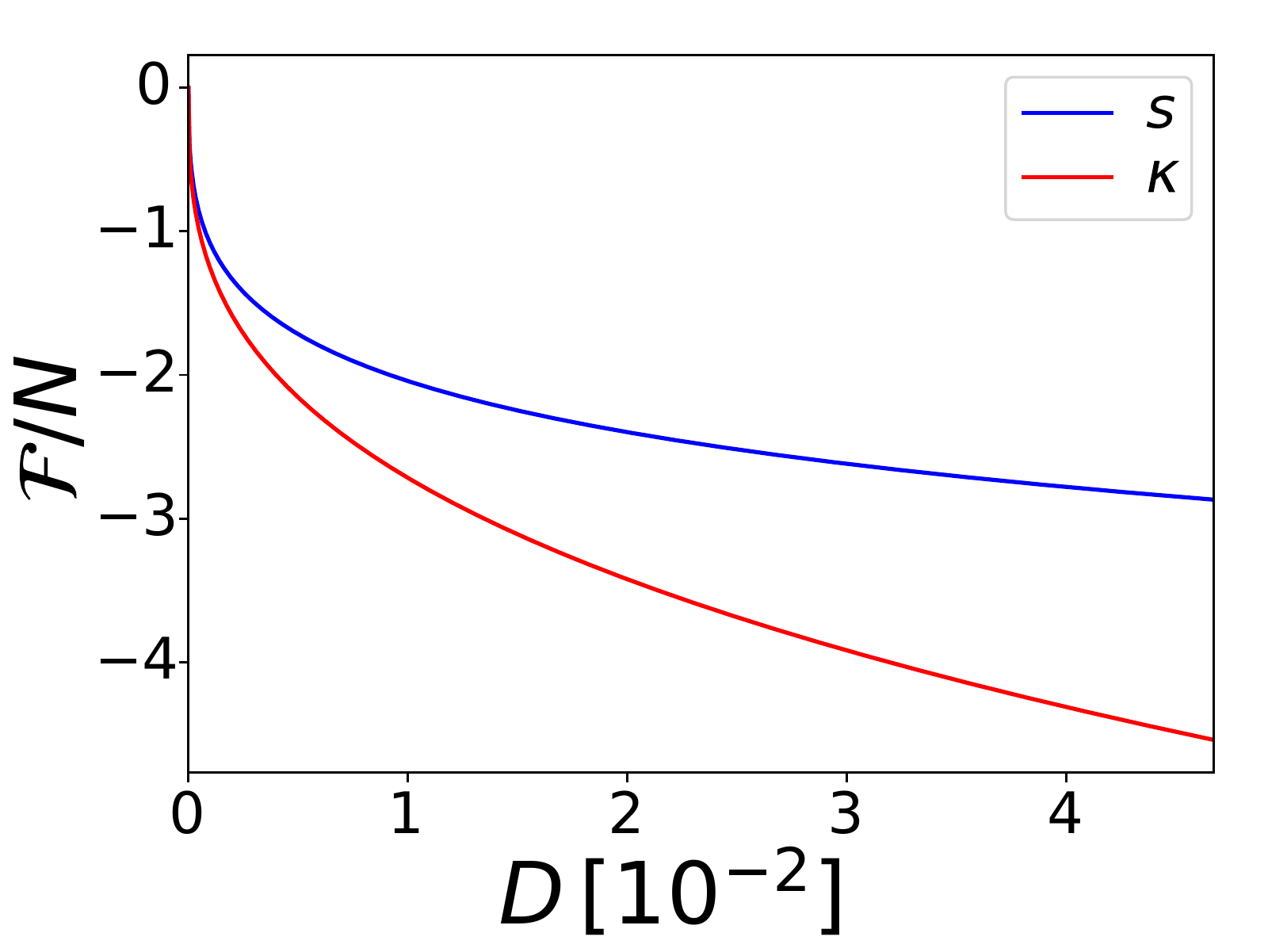}
\caption{Bogoliubov functional $\mathcal{F}$ as a function of the relative depletion $D$ along the straight path $s$ \eqref{nt} and the curved path $\kappa$. Dilute Bose gas in 3D for $n=10^{-3}$, $W_{\mathbf{0}}=m=1$ and $a_1=-0.01$ (left), charged Bose gas in 3D for $2m=e^2/2=1$ and $n=100$ (right).}
\label{fig:Fdil}
\end{figure}

The result for $\mathcal{F}$ as a function of the fraction $D$ of non-condensed bosons along the two paths $s$ and $\kappa$ is shown in Fig.~\ref{fig:Fdil} for the parameters $n=10^{-3}$, $W_{\mathbf{0}}=m=1$ and $a_1=-0.01$. This choice of parameters (recall that we set several physical constants to one) corresponds to realistic dilute Bose gases as our following results of the small degrees of depletion will confirm. We observe that the Bogoliubov functional $\mathcal{F}$ goes to zero for $D=0$ which corresponds to complete BEC. Also, it can be seen that the gradient of $\mathcal{F}$ increases for smaller distances $D$. In Sec.~\ref{sec:BECforce} we will show that the gradient of $\mathcal{F}$ actually diverges in the limit $D\to 0$ and provide a detailed discussion of this remarkable and far-reaching observation.

A generalization of $\mathcal{F}$ given by Eq.~\eqref{FBog} to dimensions $d\neq 3$ within the s-wave scattering approximation is possible if the Bogoliubov approximation for the given set of parameters, i.e.~the interaction strength and the density, is valid. Two-dimensional dilute systems are weakly interacting if the condition $n|a|_{2\mathrm{D}}^2\ll 1$ \cite{Schick71, Lieb00} is satisfied where $a_{2\mathrm{D}}$ is now the respective two-dimensional s-wave scattering length. In contrast to higher dimensional systems, a one-dimensional Bose gas is weakly interacting in the limit of high densities and the validity of the Bogoliubov approximation in that limit was shown in Ref.~\cite{Lieb63}. Due to their distinctive role, we will study a one-dimensional model in Sec.~\ref{subsec:Hubbard5}.

\subsection{Charged Bose gas in 3D}\label{subsec:charged}
In contrast to the dilute Bose gas discussed in the previous section, the scattering of \emph{charged} bosons cannot be described within the s-wave scattering approximation anymore. This is due to the infinite range of the Coulomb interaction $W(r)\propto 1/r$. The respective Fourier coefficients $W_{\bd{p}}$ can still be determined analytically though. In case of an additional uniform background they follow as
\begin{equation}\label{Wpch}
W_{\textbf{0}}=0\,,\quad W_{\bd{p}} = \frac{4\pi e^2}{p^2}\,,\,\forall \bd{p}\neq \mathbf{0}\,.
\end{equation}
For charged bosons the weak interaction regime corresponds to the high density limit \cite{Foldy61, Girardeau62, Lieb01}. This regime
to which Bogoliubov's approximation refers to is characterized by a small ``gas parameter'', $r_s \equiv (3/4\pi)^{1/3}me^2 n^{-1/3}\ll 1$. To illustrate again how RDMFT works, we calculate the energy and momentum occupation numbers $n_{\bd{p}}$ of the ground state for the most realistic case of a kinetic energy given by $\hat{t}=\sum_{\bd{p}}\frac{p^2}{2m} \hat{n}_{\bd{p}}$. For this, we add
the exact kinetic energy functional $\mbox{Tr}_1[\hat{t}(\cdot)]$ to the universal interaction functional $\mathcal{F}$ with Fourier coefficients $W_{\bd{p}}$ given by Eq.~\eqref{Wpch}. Then, we minimize the total energy functional with respect to all $\bd{n} \in \triangle$, leading to the minimizer $\thickbar{\bd{n}}$ which is given by Eq.~\eqref{nbar}. Evaluating then the functional at $\thickbar{\bd{n}}$ is straightforward (in contrast to the dilute neutral Bose gas) and one finds (recall $n \equiv N/V$)
\begin{equation}\label{Fc}
%\thickbar{\mathcal{F}} \equiv
\mathcal{F}(\thickbar{\bd{n}}) = \frac{2 \Gamma\left(-\frac{1}{4}\right)\Gamma\left(\frac{7}{4}\right)Nn^{1/4}e^{5/2}m^{1/4}}{3\pi^{5/4}}
\end{equation}
and the respective fraction of non-condensed bosons $D=1-N_\mathrm{BEC}/N$ follows as
\begin{equation}\label{Dch}
\thickbar{D}\equiv D(\thickbar{\bd{n}}) = -\frac{\Gamma\left(-\frac{3}{4}\right)\Gamma\left(\frac{5}{4}\right)m^{3/4}e^{3/2}}{4\pi^{7/4}n^{1/4}}\,.
\end{equation}
Eq.~\eqref{Dch} verifies that the depletion of the condensate decreases with increasing density $n$.
Adding the kinetic energy to Eq.~\eqref{Fc} leads to the known result for the ground state energy ($\hbar = 4\pi\epsilon_0=1$) \cite{Foldy62}:
\begin{equation}\label{Ech}
E = -\frac{4\Gamma\left[-\frac{5}{4}\right]\Gamma\left[\frac{7}{4}\right]Nn^{1/4}m^{1/4}e^{5/2}}{3\pi^{5/4}}\,.
\end{equation}
As a consistency check, this confirms the correctness of Eq.~\eqref{Fc}.

Next, in analogy to Sec.~\ref{subsec:dilute} we consider again the straight path $s$ and the curved path $\kappa$. The latter is again defined as the curve $\bd{n}(\kappa)$ obtained by reducing by  factor $\kappa \in [0,1]$ the coupling strength of the Hamiltonian above (which led to the results Eqs.~\eqref{Fc}, \eqref{Dch} and \eqref{Ech}).
Evaluating the distance $D$ along the path $\kappa$ yields
\begin{equation}
D(\kappa) = \thickbar{D}\kappa^{3/4}
\end{equation}
and the functional $\mathcal{F}$ takes the values
\begin{equation}\label{Fchkappa}
\mathcal{F}[\bd{n}(\kappa)] =
q D^{1/3}(\kappa) \,,
\end{equation}
$q \equiv 2^{5/3}\Gamma\left(-\frac{1}{4}\right)\Gamma\left(\frac{7}{4}\right)Nn^{1/3} e^2/3\left[-\Gamma\left(-\frac{3}{4}\right)\Gamma\left(\frac{5}{4}\right)\right]^{1/3}\!\pi^{2/3}$.
For the  path $s$ we have $D(t) = (1-t)\thickbar{D}$ and the concrete values of the functional $\mathcal{F}$ along that path can be evaluated by exact numerical means. The right panel of Fig.~\ref{fig:Fdil} shows $\mathcal{F}$ along the two paths $s$ and $\kappa$. The curves have qualitatively similar shapes to those for the dilute neutral Bose gas shown on the left of Fig.~\ref{fig:Fdil}. This is not surprising because both setups correspond to a weakly interacting system in which the Bogoliubov approximated functional Eq.~\eqref{FBog} is valid. However, we will see in Sec.~\ref{sec:BECforce} that the momentum dependence of $W_{\bd{p}}$ can alter the behaviour of the gradient of $\mathcal{F}$.

\subsection{Bose-Hubbard model for five lattice sites}\label{subsec:Hubbard5}
As a third example, we discuss in this section the one-dimensional Bose-Hubbard model.
For illustrative purposes, we consider the specific case of just $L=5$ lattice sites and $N=100$ bosons since this allows us to visualize
the functional and its gradient on the entire domain $\triangle$. Indeed, for $L=5$ there are only two independent momentum occupation numbers due to the general parity symmetry $n_{\bd{p}}= n_{-\bd{p}}$ and normalization $n_{\bd{0}}= N-\sum_{\bd{p} \neq 0} n_{\bd{p}}$.

We start by discussing a few conceptual aspects which are valid for any number $L$ of sites (assuming for simplicity $L$ odd). The one-dimensional Brillouin zone comprises momenta $ p= 2\pi\nu/L$ where $\nu$ takes integer values in the interval described by $|\nu| \leq  (L-1)/2$.
The bosons interact via Bose-Hubbard on-site interaction as described by the operator $\frac{U}{2}\sum_{j=1}^L\hat{n}_j\left(\hat{n}_j-1\right)$. It is worth recalling that the universal functional depends on the entire interaction Hamiltonian, i.e., it includes \emph{a priori} the coupling constant $U$ as well. Yet, due to the linear structure of the constrained search \eqref{Levy}, \eqref{LevyP} any non-negative prefactor could be separated from the interaction Hamiltonian $\hat{W}$ and added instead in front of the respective universal functional,
\begin{equation}
\F_{U \hat{W}} = |U| \, \F_{ \mathrm{sgn}(U)\hat{W}}\,.
\end{equation}
It is crucial to observe that the same does not apply to possible sign factors since otherwise this would mean to change the minimization in \eqref{Levy}, \eqref{LevyP} to a maximization. Because of this, we consider in the following the interaction Hamiltonian $\hat{W} \equiv \frac{\mathrm{sgn}(U)}{2}\sum_{j=1}^L\hat{n}_j\left(\hat{n}_j-1\right)$ and add eventually the coupling constant $|U|$
in front of the respective universal functional $\F_{\hat{W}}$. To proceed, it is then an elementary exercise to determine the corresponding Fourier coefficients $W_p=\mathrm{sgn}(U)$ which are in particular independent of the (one-dimensional) momentum $p$. The universal functional in the BEC regime is obtained by plugging the concrete result for the Fourier coefficients $W_p$ into the general formula for the Bogoliubov functional \eqref{FBog} or its extension \eqref{Fmin} based on Girardeau's approach. Just to reiterate, the respective functionals are valid in the regime of BEC, i.e., for weak interactions. In contrast to their higher dimensional counterparts, one-dimensional systems require high densities $n=N/L \gg 1$ to be weakly interacting \cite{Lieb63}.

From a general point of view, the context of lattice models emphasizes very well the conceptual advantages of RDMFT relative to wavefunction based methods. After having determined the universal interaction functional $\F_{\hat{W}}$ (or decent approximations thereof) the ground state energy of every Hamiltonian $\hat{H}(\hat{t}) = \hat{t}+ |U|\,\hat{W}$ can be calculated with relatively little computational effort by minimizing the total energy functional with respect to all $\bd{n} \in \triangle$. In that sense, RDMFT represents a highly economic approach for solving simultaneously the ground state problem for the entire class $\{\hat{H}(\hat{t})\}$ of Hamiltonians. For continuous systems the benefits of this are less obvious since there is essentially one particularly relevant choice for the kinetic energy operator $\hat{t}$. This is quite different for lattice models since both
the rate and the range of the hopping can be varied in experiments (see, e.g., Refs.~\cite{Guenter2013,Schempp2015}). Nonetheless, we focus in the following on hopping just between neighbouring sites at a rate $t\geq 0$, i.e., we choose $\hat{t} = -2t\sum_{p}\big(\!\cos(p)-1\big)\hat{n}_p$ and w.l.o.g.~fix $|U|\equiv 1$. In analogy to the most realistic dilute and charged Bose gas as discussed in Sec.~\ref{subsec:dilute} and Sec.~\ref{subsec:charged}, respectively, we pick $t=U=1$ as a reference point for further investigations and illustrations .
\begin{figure}[htb]
\includegraphics[width=0.6\linewidth]{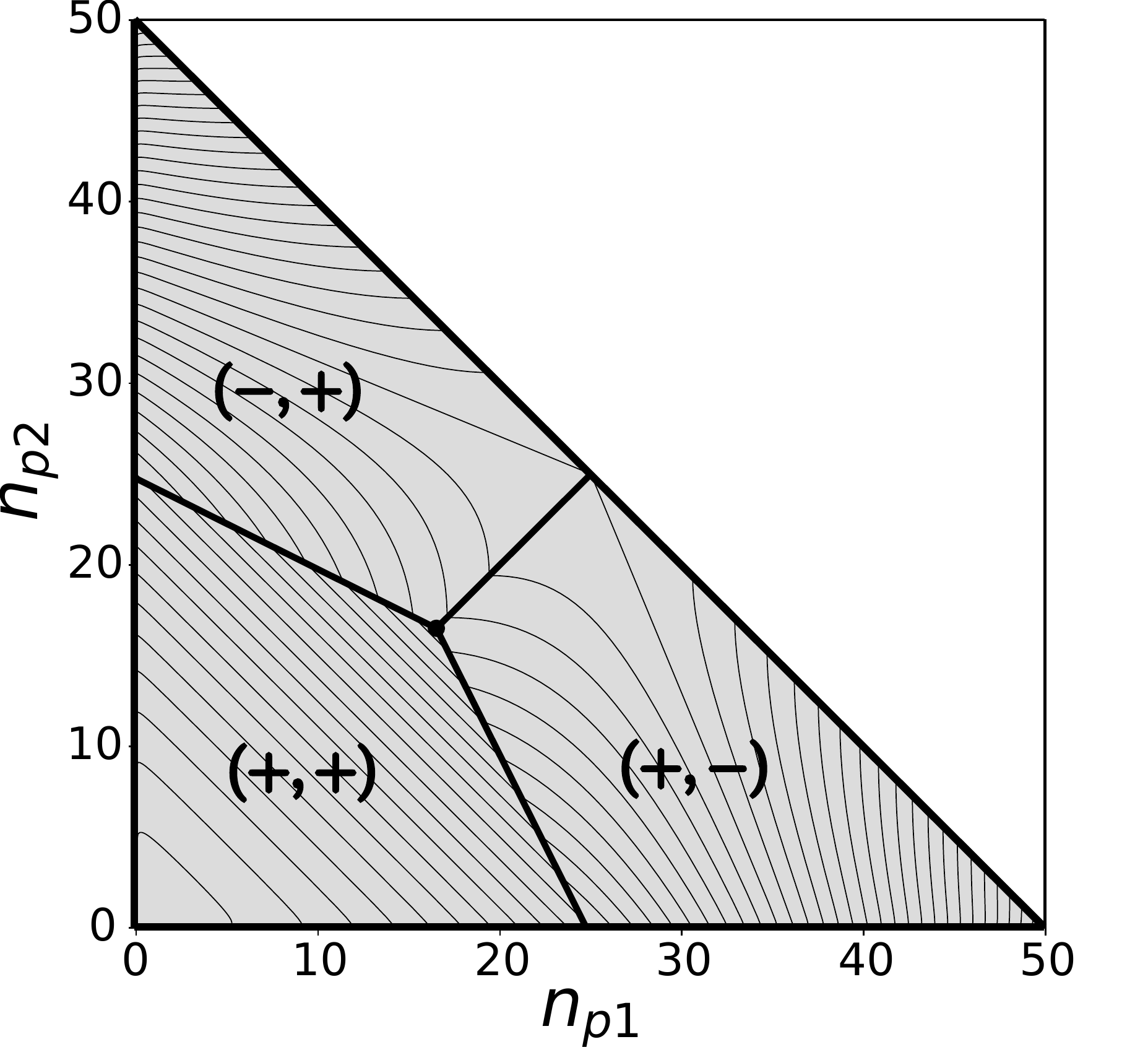}
\caption{Domain $\triangle$ of the universal functional is shown for $L=5$ sites.
Minimization of signs $(\sigma_{p_1},\sigma_{p_2})$ in \eqref{Fmin} partitions $\triangle$ into three cells (see text for details).
\label{fig:S}}
\end{figure}

To illustrate and compare the Bogoliubov- \eqref{FBog} and the Girardeau-approximated functionals \eqref{Fmin} for the specific case of $L=5$ sites we first need to execute the minimization of the sign factors in \eqref{Fmin}. As it is shown in Appendix \ref{app:signs}, this can be done analytically due to the specific Fourier coefficients and leads to a partitioning of the functional's domain $\triangle$ into three regions. Just for illustrative purposes, we present in Fig.~\ref{fig:S} the \emph{entire} domain $\triangle$ of the functional \eqref{Fmin} (recall its validity refers to the regime of BEC only) and the three cells which are characterized by different  minimizing sign configurations $(\sigma_{p_1},\sigma_{p_2})$ in \eqref{Fmin}. As the two independent occupation numbers we choose here the momenta $p_1=2 \pi/5$ and $p_2=4 \pi/5$ which can take values $n_{p_j} \in [0,50]$. The vector $(0,0)$ corresponds to complete BEC, i.e., $N_{\mathrm{BEC}}\equiv n_{0}=N=100$  and its vicinity represents the BEC-regime to which our functionals refer.
\begin{figure}[htb]
\includegraphics[width=0.43\linewidth]{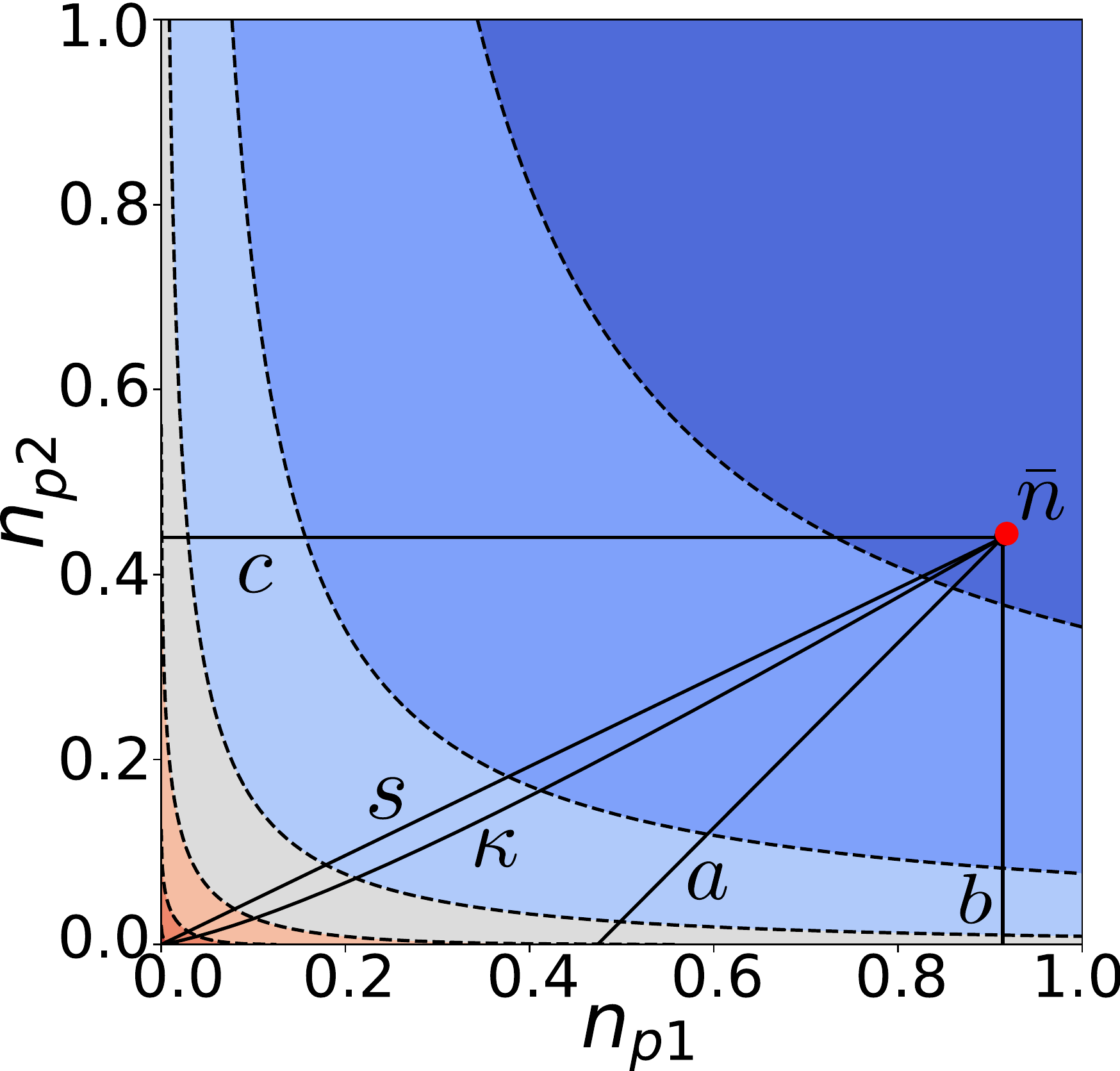}
\includegraphics[width=0.43\linewidth]{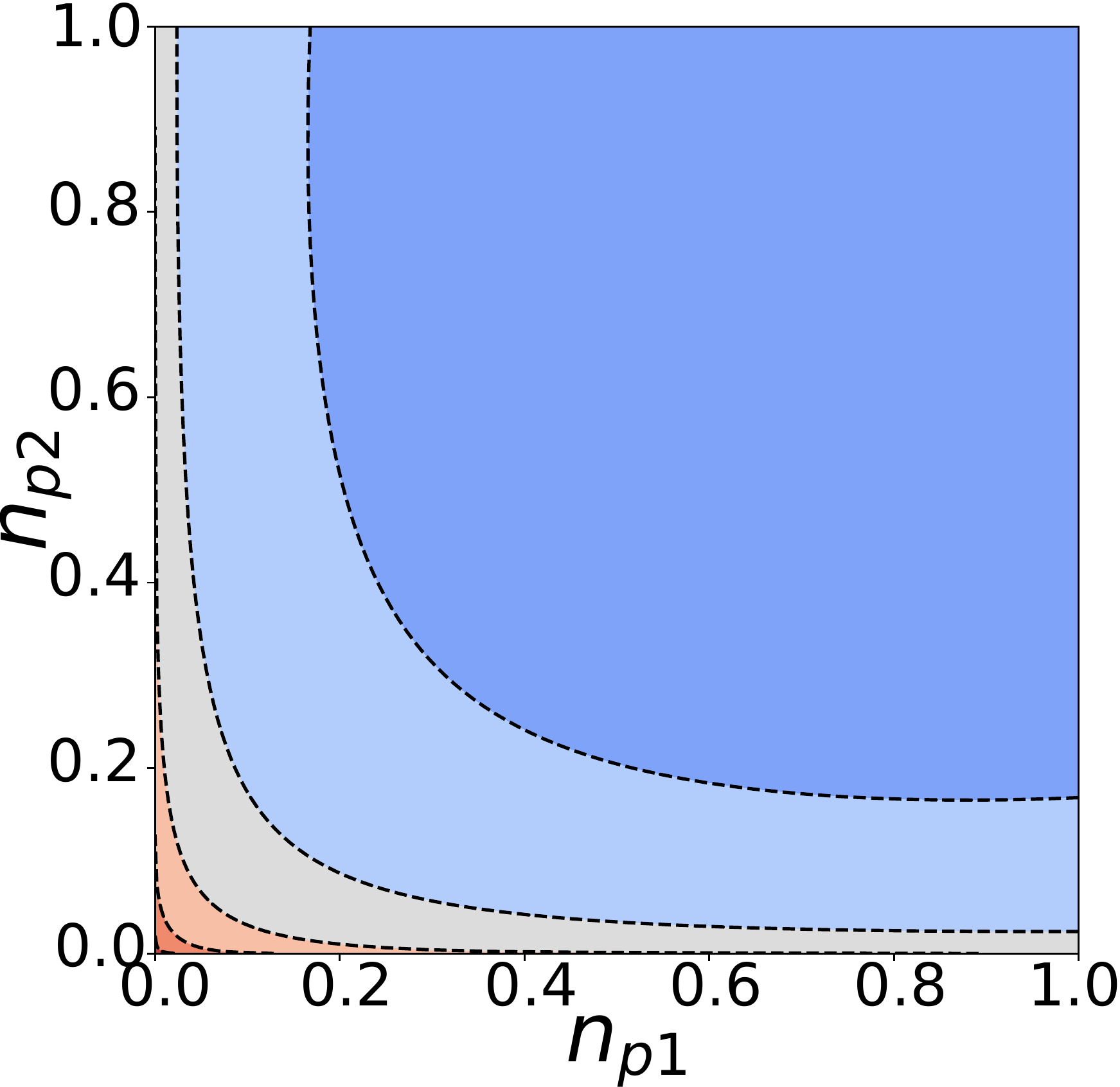}
\hspace{0.1cm}
\includegraphics[width=0.07\linewidth, trim=0cm -1.5cm 0cm +1.5cm]{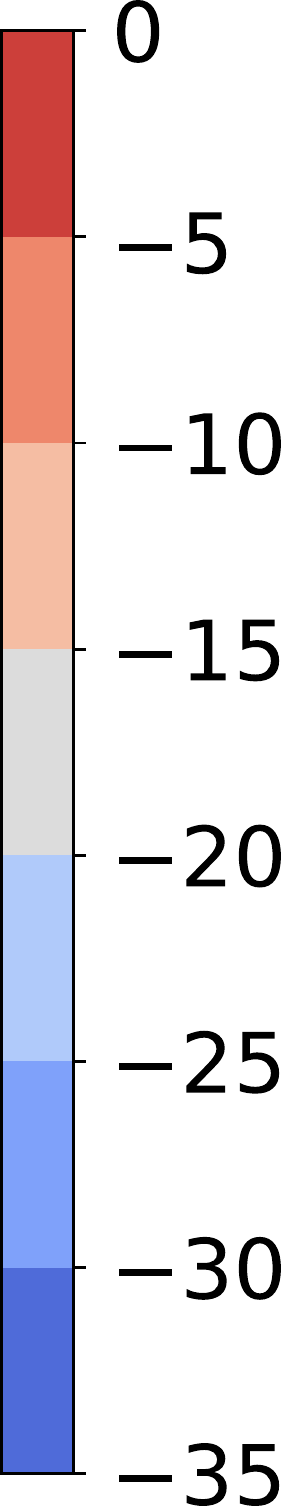}
\caption{Left: Contour plot of the Bogoliubov-approximated functional \eqref{FBog} for the Bose-Hubbard model with $N=100$ bosons on $L=5$ sites in the BEC-regime of not too large depletion. Right: Girardeau's extension \eqref{Fmin} for the same system.}
\label{fig:HubbardF}
\end{figure}

In Fig.~\ref{fig:HubbardF} we present now the Bogoliubov functional \eqref{FBog} and its extension \eqref{Fmin} based on Girardeau's approach
in the form of a contour plot in the BEC-regime of not too large quantum depletion.
The results for the two functionals are in quite good agreement for small degrees $D$ of depletion. The occupation number vector $\thickbar{n}=(0.91, 0.44)$  obtained from minimizing the total energy functional for the reference point $(t,U)=(1,1)$ is shown in Fig.~\ref{fig:HubbardF} as well. The corresponding degree of depletion, $D=2.7\%$, justifies in retrospective the treatment of the interaction $\hat{W}$ within the Bogoliubov theory and the usage of the functionals \eqref{FBog} and \eqref{Fmin}, respectively. For stronger quantum depletion the two functionals in Fig.~\ref{fig:HubbardF} begin to differ also qualitatively. Their (small) deviation already in the regime of BEC with a degree of depletion around $2\%$ emphasizes the quantitative significance of the additional terms $I_1$ and $I_2$ and the usage of the exact value $n_{\mathbf{0}}$ rather than its approximation to $N$ in Eq.~\eqref{Fmin}.

\begin{figure}[htb]
\includegraphics[width=0.8\linewidth]{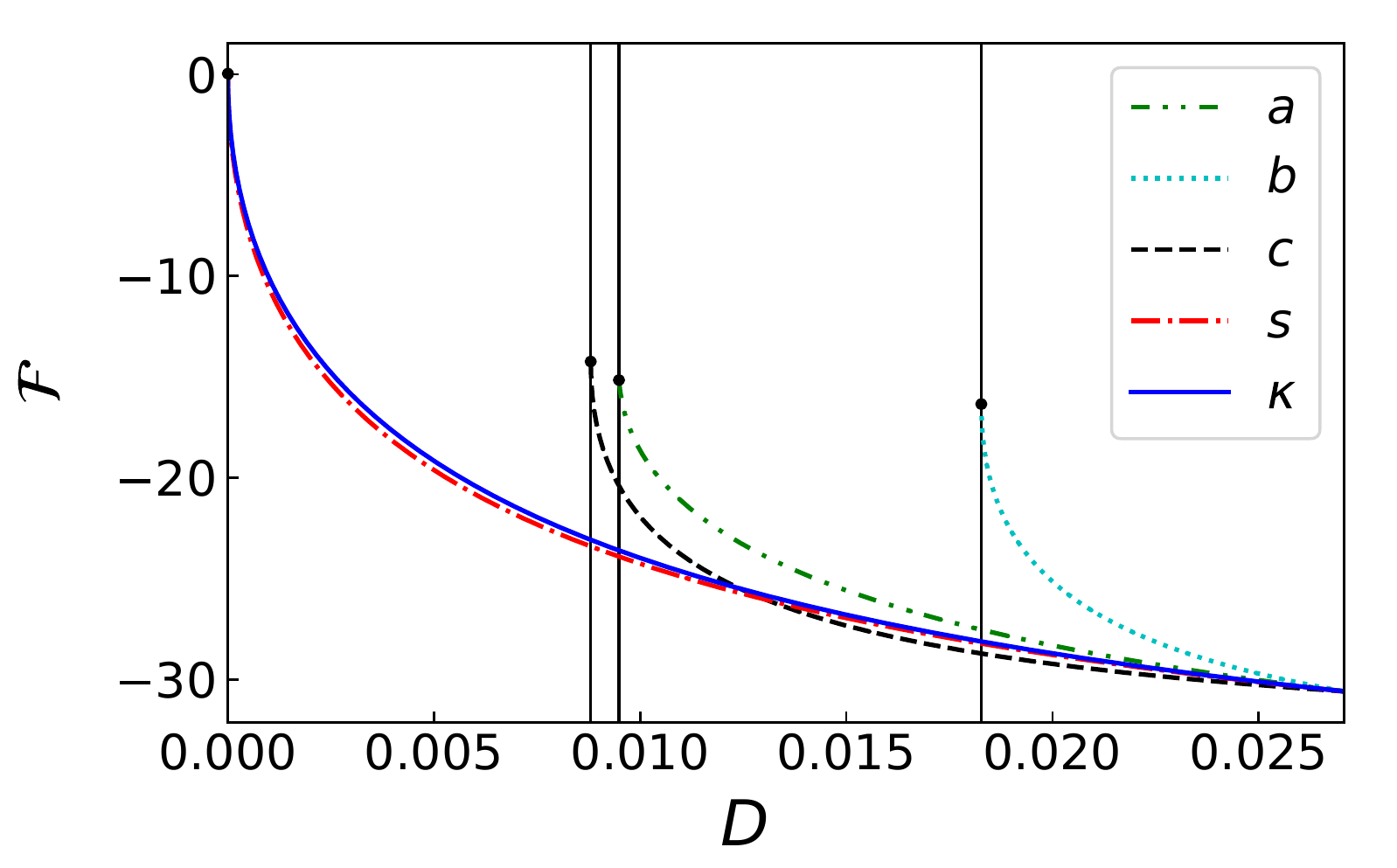}
\caption{Universal functional $\mathcal{F}$ for the Bose-Hubbard model along the five paths defined in Fig.~\ref{fig:HubbardF}}
\label{fig:FD}
\end{figure}

For the discussion in Sec.~\ref{sec:BECforce} of the new concept of a BEC force, we define in Fig.~\ref{fig:HubbardF} five qualitatively different paths towards the polytope boundary $\partial\triangle$, all starting from the point $\thickbar{n}$. The path denoted by $s$ corresponds to a straight path towards complete BEC and $\kappa$ denotes the path where the interaction strength of the model is reduced by increasing the kinetic energy by a factor $1/\kappa$ with $\kappa \in[0,1]$. The path denoted by $a$ runs perpendicular towards the hyperplane defined by $n_{p_1} + n_{p_2}=0$. Consequently, it corresponds to the path with the fastest increase of the condensate fraction (yet it will not reach complete BEC). In the cases $b$ and $c$ one occupation number is fixed while the other one is continuously decreased to zero. In Fig.~\ref{fig:FD} we present the functional $\mathcal{F}$ as a function of $D=1-N_\mathrm{BEC}/N$ along those five paths. The black dots emphasize that the value of $\mathcal{F}$ at the boundary $\partial\triangle$ remains finite (quite in contrast to its derivative as shown and discussed in the subsequent section).
The convexity of the curves in Fig.~\ref{fig:FD} corresponding to the four straight paths $a, b, c, s$ just reflects the local convexity of the exact universal functional in the regime of not too large quantum depletion.
%(in which it coincides with the convex Bogoliubov functional).
In this context, we would like to reiterate that this convex behavior and the repulsive character of the functional's gradient close to the boundary emerges from the minimization of the sign factors in \eqref{Fmin} leading to \eqref{sgnmagic}.

\section{Repulsive Bose-Einstein condensation force}\label{sec:BECforce}
In this section we explore in more detail the behavior of the functional and its gradient close to the boundary of their domain
$\triangle$ in the regime of BEC. This will eventually allow us to reveal and establish the novel concept of a BEC force.  Due to its potentially far-reaching consequences for our understanding of bosonic quantum systems, we will calculate and illustrate the BEC force in Sec.~\ref{subsec:forceex} for the three different systems studied in Sec.~\ref{sec:appl}.

\subsection{General results}\label{subsec:forcegeneral}
The functional \eqref{FBog} which is based on the Bogoliubov approximation is convex on its entire domain $\triangle$ \eqref{simplex}. Since this approximate functional is exact in leading order in the regime of BEC with not too large quantum depletion all conclusions drawn from it are valid for the exact universal functional of \eqref{H1st} as well. The illustrations in the previous section for dilute and charged Bose gases in 3D and the Bose-Hubbard model have also confirmed the distinctive convex behavior of the universal functional in the BEC-regime. While the functional itself remains finite even at the point $\mathbf{0} \in \triangle$ of complete condensation, the same will not be true anymore for the functional's gradient. This can easily be deduced from the form of the Bogoliubov functional \eqref{FBog}. To be more specific, approaching the vertex $\mathbf{0}$ of the simplex  $\triangle$ means to simultaneously send all momentum occupation numbers $n_{\bd{p}}$ with $\bd{p}\neq \mathbf{0}$ to zero. Taking then the derivative of \eqref{FBog} (or of its extension \eqref{Fmin}) with respect to $n_{\bd{p}}$ sufficiently close to $\mathbf{0}$ yields in leading order
\begin{equation}\label{npforce}
\frac{\partial \F}{\partial n_{\bd{p}}}(\bd{n}) \sim - \frac{n |W_{\bd{p}}|}{2} \frac{1}{\sqrt{n_{\bd{p}}}}\,.
\end{equation}
It is worth noticing that the divergence of this derivative for $ n_{\bd{p}} \rightarrow 0$ is always repulsive for \emph{any} interaction $\hat{W}$.
This remarkable feature follows directly from the minimization of the sign factors in \eqref{Fmin}, leading to \eqref{sgnmagic}. The repulsive nature of the diverging gradient also proves universally that occupation numbers in interacting bosonic quantum systems can never attain the exact mathematical value $0$. Although our work refers to homogeneous systems in their BEC regime only, we have little doubt that this conclusion is also valid for any generic nonhomogeneous interacting bosonic quantum system, also beyond the BEC regime. For the sake of clarity, we would like to emphasize that these statements do not hold for \emph{non-interacting} bosons ($W_p=0$) since the corresponding universal functional is zero, $\F\equiv 0$.

The general result \eqref{npforce} implies that for interacting bosons the point $\mathbf{0}$ of complete condensation can never be reached, independent of the path towards $\mathbf{0}$ that is envisaged. Since the functional $\F$ is finite this seems to be paradoxical as far as the energy is concerned. Yet, the reader shall note that it is the kinetic energy which will need to diverge according \eqref{varE0} to enforce such a path towards $\mathbf{0}$.

We also would like to emphasize that the divergence of the gradient of $\F$ along a straight path is always proportional to $1/\sqrt{D}$ and its prefactor depends on the direction of the path, i.e., the angle at which $\mathbf{0}$ is approached. To confirm this in a quantitative way, let us consider a general straight path from a starting point $\thickbar{n}$ in the regime of BEC to $\mathbf{0}$, linearly parameterized by $t \in [0,1]$,
\begin{equation}\label{nstraight}
\bd{n}(t) = (1-t)\, \thickbar{\bd{n}}\,.
\end{equation}
The degree $D$ \eqref{Dgen} of quantum depletion along that path reduces according to
\begin{equation}\label{Dstraight}
D(t) = (1-t)\, \thickbar{D} \equiv (1-t)\,\frac{1}{N}\sum_{\bd{p}\neq \mathbf{0}} \thickbar{n}_{\bd{p}}\,.
\end{equation}
The gradient of $\F$ projected onto that path is then nothing else than the weighted sum of individual contributions \eqref{npforce} from every $\bd{p}$,
\begin{eqnarray}\label{BECforce}
\frac{\partial \F}{\partial D}\Big\vert_{\mathrm{path}}  &=& \nabla_{\!\bd{n}}\F \cdot \frac{\partial \bd{n}}{\partial D}\Big\vert_{\mathrm{path}} \\
&=& \nabla_{\!\bd{n}}\F \cdot \frac{\thickbar{\bd{n}}}{\thickbar{D}}
\sim -\frac{n}{2} \sum_{\bd{p}\neq \mathbf{0}} |W_{\bd{p}}|\sqrt{\frac{\thickbar{n}_p}{\thickbar{D}}}\,  \frac{1}{\sqrt{D}}\,. \nonumber
\end{eqnarray}
This second key result of our work establishes the new concept of a BEC force which prevents interacting bosonic quantum systems from ever exhibiting complete BEC. This novel concept is conceptually very similar to the fermionic exchange force that we have recently revealed and established in fermionic lattice models \cite{Schilling2019}.

\subsection{Examples}\label{subsec:forceex}
In this section we illustrate the novel concept of a BEC force \eqref{BECforce} for various systems introduced in Sec.~\ref{sec:appl}.

\subsubsection{Bose gases in 3D}
We revisit the 3D Bose gas for neutral atoms in the low density and for charged atoms in the high density regime. The aim is to calculate for those concrete systems the explicit values of the BEC force \eqref{BECforce}. For both systems, the derivative of $\mathcal{F}$ with respect to the degree $D$ of quantum depletion along the path $s$ defined by \eqref{nstraight} is given by Eq.~\eqref{BECforce}.
The summation over $\bd{p}\neq 0$ can be converted into an integral in the thermodynamic limit where $N\to \infty$, $V\to \infty$ and $n=N/V=\mathrm{cst.}$. This eventually allows us (see Appendix \ref{app:forcedil}) to obtain a compact analytic expression  for the BEC force,
\begin{equation}\label{DFsdilute}
\left.\frac{\rmd \mathcal{F}(\thickbar{\bd{n}})}{\rmd D}\right\vert_s\sim
N \left[\eta(a_0, n, m) + \frac{2\pi n a_1}{m\sqrt{\thickbar{D}}}\right]\frac{1}{\sqrt{D}}
\end{equation}
where $\eta(a_0, n, m)$ is a positive constant and $a_0$ and $a_1$ are the first two terms in the Born series for the scattering length $a$.
It is worth reiterating that according to the general result \eqref{BECforce} the BEC force is always repulsive. Since
only the second term in Eq.~\eqref{DFsdilute} is negative (recall $a_1<0$) this imposes in turn a bound on the maximal valid distance $\thickbar{D}$ of the starting point $\thickbar{n} \in \triangle$ to the regime of complete BEC.

For the charged Bose gas the last expression in the second line in Eq.~\eqref{BECforce} can only be calculated by exact numerical means. Nonetheless, this also allows us to confirm the square root dependence of the divergence. In general, the functional's gradient diverges as $1/\sqrt{\mbox{dist}(\bd{n},\partial \triangle )}$ along straight paths reaching any \emph{arbitrary} point on the boundary $\partial \triangle$ in the regime of BEC.

\begin{figure}[htb]
\includegraphics[width=0.48\linewidth]{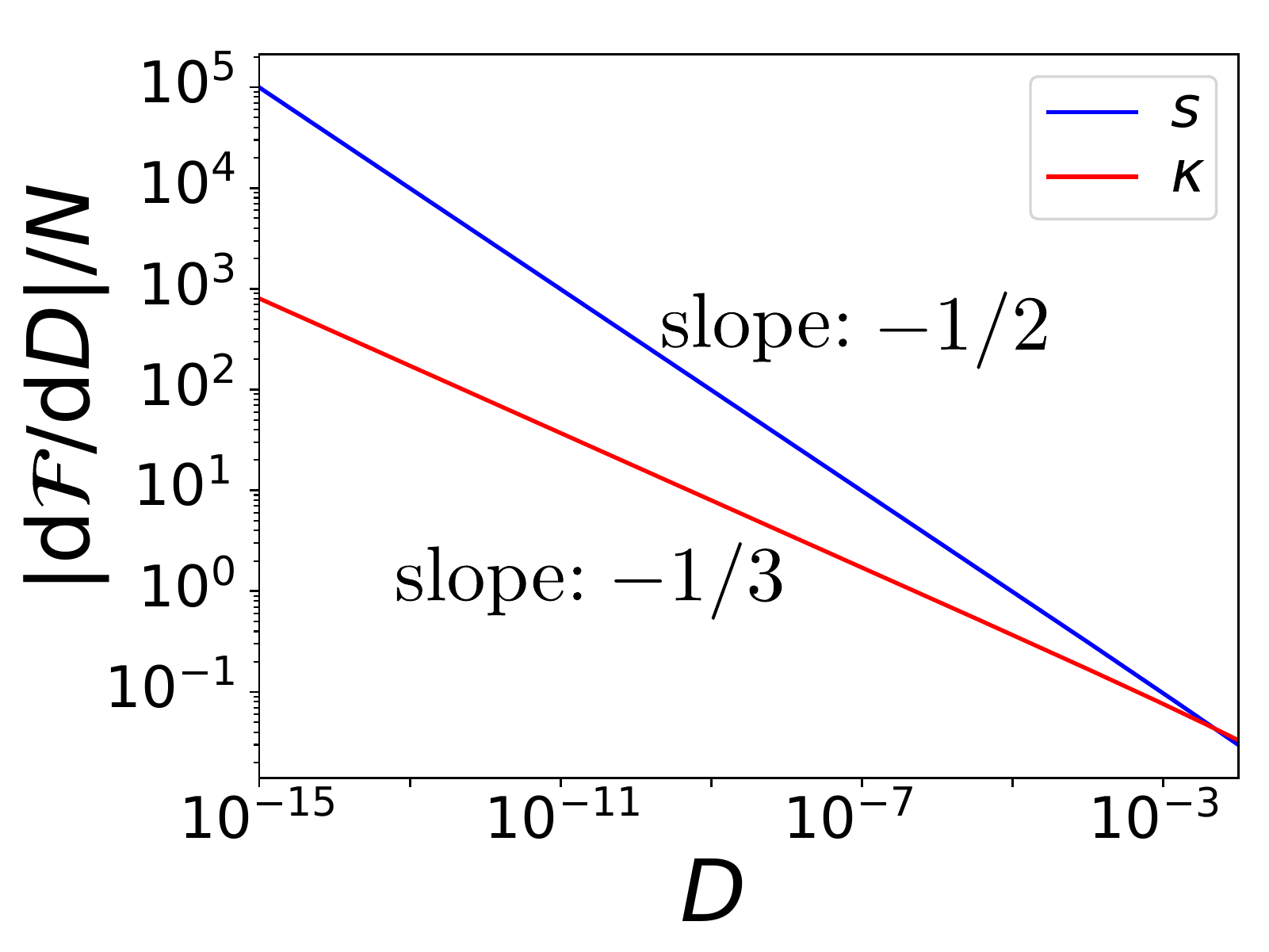}
\hspace{0.1cm}
\includegraphics[width=0.48\linewidth]{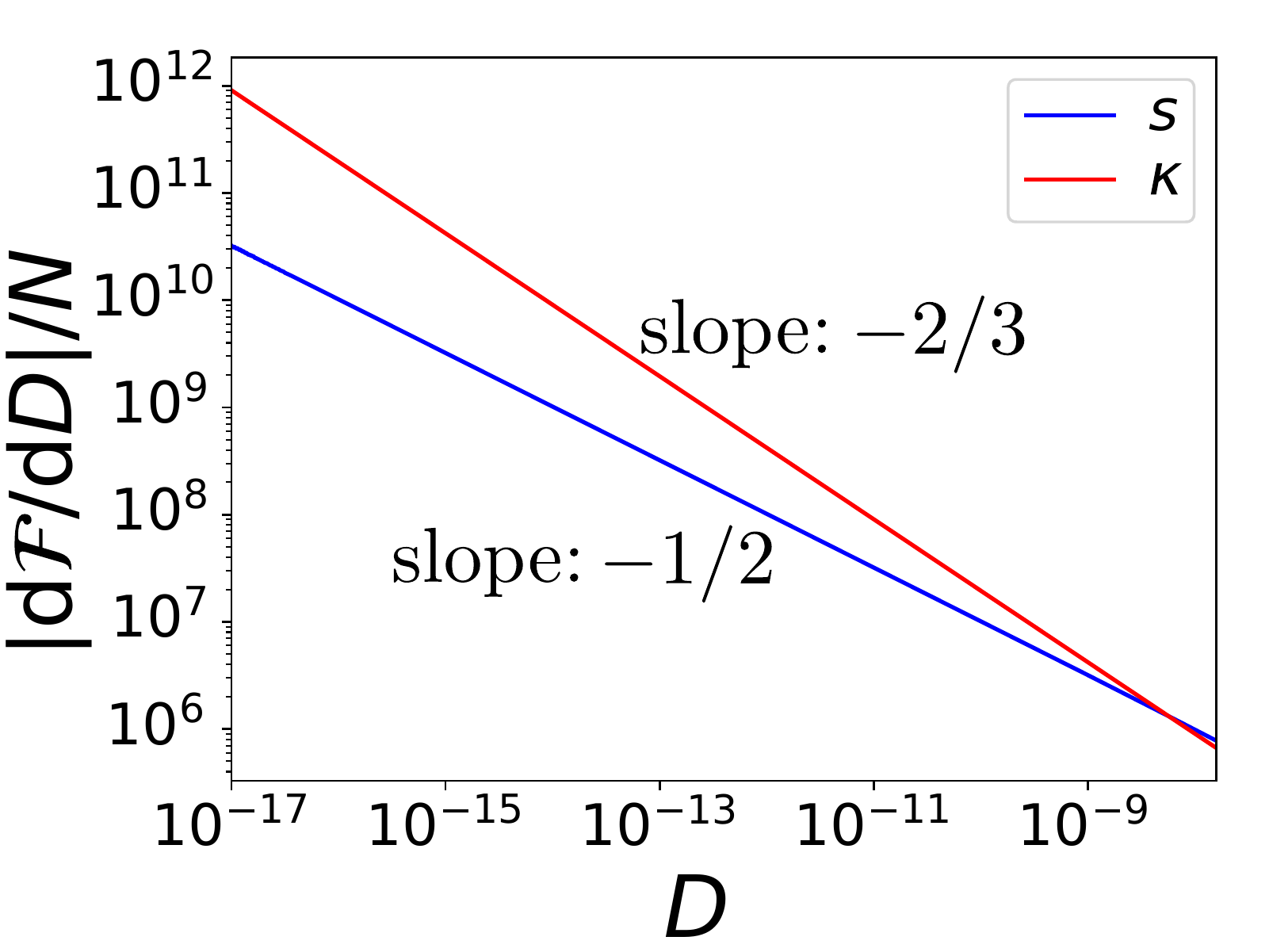}
\caption{BEC force $|\rmd \mathcal{F}/\rmd D|$ along the straight path $s$ (blue) and the curved path $\kappa$ (red) is shown for the dilute Bose gas in 3D with $n=10^{-3}$, $W_{\mathbf{0}}=m=1$ and $a_1=-0.01$ (left) and for the charged Bose gas in 3D with $2m=e^2/2=1$ and $n=100$ (right).
\label{fig:dF}}
\end{figure}

Moreover, we determine for both systems the BEC force along the curved path $\bd{n}(\kappa)$ which is defined by reducing an additional coupling constant $\kappa$ in front of $\hat{W}$ from one to zero. Since exactly this path has been implemented in a very recent experiment \cite{Lopes17} this may suggest a first experimental setup for realizing and visualizing our novel concept of a BEC force: The closer $\bd{n}(\kappa)$ is to the point $\bd{0}$ of complete condensation, the more difficult it will get to further approach that point. This will manifest itself in more and more suppressed responses of the BEC to external perturbations.
The explicit calculation of the BEC force along the $\kappa$-path follows directly from differentiation of the expressions in \eqref{Fdilkappa} and \eqref{Fchkappa}, respectively, leading to
\begin{equation}\label{dFdilkappa}
\left.\frac{\rmd \mathcal{F}_\mathrm{dilute}}{\rmd D}\right\vert_\kappa\propto -\frac{1}{D^{1/3}}
\end{equation}
and
\begin{equation}\label{dFchkappa}
\left.\frac{\rmd \mathcal{F}_\mathrm{charged}}{\rmd D}\right\vert_\kappa\propto -\frac{1}{D^{2/3}}\,.
\end{equation}
Fig.~\ref{fig:dF} displays the BEC force along the straight $s$-path and the curved $\kappa$-path for both ultracold gas systems. The linear behavior shown in this log-log plot confirms the algebraic dependence of the BEC force on the degree $D$ of quantum depletion along both paths.
For the dilute Bose gas, the gradient of $\mathcal{F}$ according to Eq.~\eqref{DFsdilute} and Eq.~\eqref{dFdilkappa} diverges faster along the path $s$ than along the path $\kappa$. For the charged Bose gas we observe the opposite behaviour.
%Indeed, the numerical result verifies $\rmd \mathcal{F}/\rmd D|_s \propto -1/\sqrt{D}$.

\subsubsection{Bose-Hubbard model}
\begin{figure}[htb]
\includegraphics[width=0.8\linewidth]{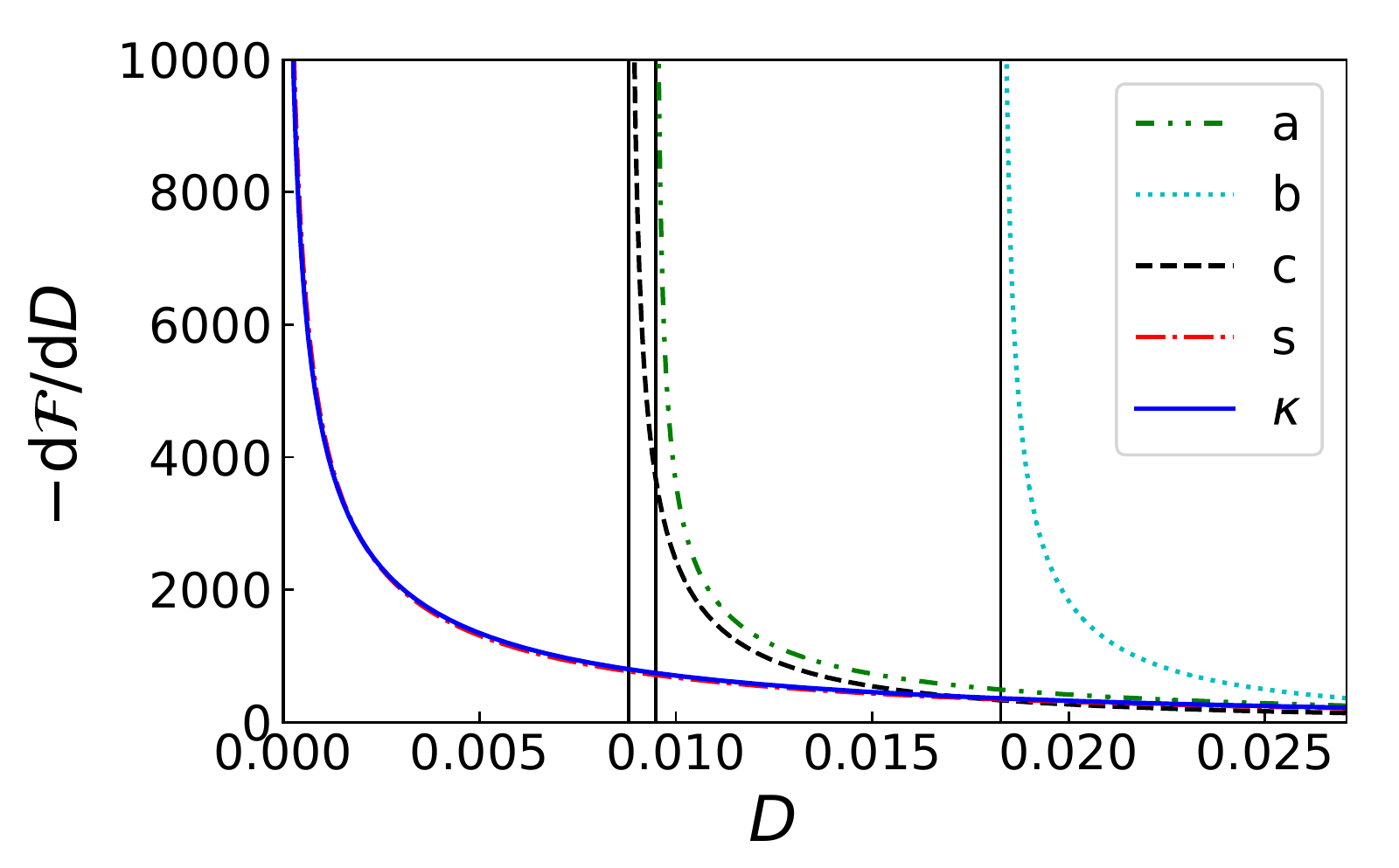}
\caption{Gradient of the universal functional $\mathcal{F}$ for the Bose-Hubbard model along the five paths defined in Fig.~\ref{fig:HubbardF}. The results for $\kappa$ and $s$ almost coincide.
\label{fig:dFD}}
\end{figure}

We illustrate the BEC force and the diverging behaviour of the functional's gradient close to the boundary $\partial \triangle$ of its domain in general for the Bose-Hubbard model. For this we consider again as in Sec.~\ref{subsec:Hubbard5} the case of $N=100$ bosons on $L=5$ sites. We then determine the directional derivative of the functional along the five paths which were defined in Fig.~\ref{fig:HubbardF}. Since for all five paths the distance $D$ of the occupation number vector $\bd{n}$ to $\mathbf{0}$ is monotonously decreasing we can parameterize the functional's derivative along each path by $D$. The respective results are depicted in Fig.~\ref{fig:dFD}. There, the vertical solid lines correspond to the values of $D$ at which the respective paths reach the boundary of $\triangle$ (see also Figs.~\ref{fig:HubbardF}, \ref{fig:FD}).
We first observe that for all five paths $-\partial \F/\partial D$ is diverging at the end point of each path on the boundary $\partial \triangle$. As a rather elementary analysis reveals (not shown here) this divergence is always proportional to $1/\sqrt{\mbox{dist}(\bd{n},\partial \triangle)}$. As far as the four straight paths $a, b, c, s$ are concerned, this was expected given the general results of Sec.~\ref{subsec:forcegeneral}. In contrast to the two continuous Bose gases, however, the same applies in the Bose-Hubbard model also for the curved $\kappa$-path which is obtained by just reducing the coupling strength.

\section{Summary and Conclusion}\label{sec:concl}
For homogeneous Bose-Einstein condensates (BECs) which are described by the Bogoliubov theory we succeeded in deriving the universal interaction functional $\F(\bd{n})$. Our approach based on the constrained search formalism necessitated, however, a distinctive
particle-number conserving modification of the conventional Bogoliubov theory.
The crucial ingredient of our derivation has been the strong connection \eqref{np} between the vectors $\bd{n}$ of momentum occupation numbers and the Bogoliubov trial states. This has drastically simplified the Levy-Lieb constrained search and the minimization \eqref{Fmin} with respect to the remaining independent degrees of freedom (sign factors) could be executed as well, eventually leading to our key result \eqref{FBog}. The universal functional $\F(\bd{n})$ comprises separate contributions of various momentum pairs $(\bd{p},-\bd{p})$ (recall $n_{\bd{p}}=n_{-\bd{p}}$). This of course resembles the fact that Bogoliubov theory describes \emph{independent} pair excitations of condensed bosons, $(\bd{0},\bd{0})\mapsto (\bd{p},-\bd{p})$.

While typical experimental realizations of BEC involve ultracold gases in their dilute regime, it is worth recalling that the scope of Bogoliubov theory is wider. As a matter of fact, it applies to any homogeneous system in arbitrary dimensionality as long as it exhibits BEC with not too large depletion. Thus, to identify the range of applicability of our derived first-level bosonic RDMFT functional one just needs to understand the conditions under which different physical systems exhibit BEC. For instance, for neutral bosons in dimensions $d\geq 2$ this would be the dilute regime (which implies effectively a weak coupling) while for $d=1$ and also for charged bosons in $d=3$ it would be the high density regime. It will be one of the future challenges to determine also the functional (or at least approximations thereof) in other physical regimes beyond BEC. Similarly to the Hartree-Fock functional in fermionic RDMFT \cite{LiebHF} and the local density approximation in density functional theory \cite{GD95}, the Bogoliubov functional could then serve as a starting point for such approximations. That's particularly promising, since in contrast to the Hartree-Fock functional our functional already involves some quantum correlations while the former always leads to occupation numbers identical to just one and zero \cite{LiebHF}.

The second key result of our work is the emergence of a universal BEC force which prevents general quantum systems of interacting bosons from ever reaching complete BEC. To be more specific, the gradient of the universal functional has been found to repulsively diverge as $1/\sqrt{1-N_{\mathrm{BEC}}/N}$ in the regime of almost complete BEC. This BEC force can be seen as a collective force since it comprises individual diverging terms $-\partial \F/\partial n_{\bd{p}} \sim n |W_{\bd{p}}|/2 \sqrt{n_{\bd{p}}}$ from each momentum $\bd{p}$.
It is worth emphasizing that it has been exactly the minimization \eqref{Fmin} of the phase factors within the constrained search formalism which eventually made this BEC force and more generally $\nabla_{\bd{n}}\F$ close to the boundary of its domain $\triangle$ \emph{repulsive} regardless of the signs of various Fourier coefficients $W_{\bd{p}}$. The BEC force provides an alternative explanation for quantum depletion which is most fundamental in the sense that it is merely based on the geometry of density matrices and the properties of the partial trace. Indeed, the type of interaction between the bosons only affects the prefactor of the BEC force, while its diverging behaviour proportional to $1/\sqrt{1-N_{\mathrm{BEC}}/N}$ is universal. Due to this universal behaviour one may expect that the respective prefactor could provide important insights into system-specific properties, in some analogy to the so-called Tan's contact \cite{Tan1,Tan2,Tan3}.
Moreover, the BEC force is nothing else than the bosonic analogue of the recently discovered fermionic exchange force \cite{Schilling2019}, showing exactly the same diverging behaviour close to the boundary of the fermionic functional's domain.

Last but not least, we comment on important follow-up challenges besides the possible experimental realization of the BEC force in ultracold gases and the construction of functionals with a larger range of validity. One such direction would be to quantify in a \emph{mathematically rigorous} way the deviation of the Bogoliubov functional from the exact functional $\F_{\hat{W}}$ of the full interaction Hamiltonian $\hat{W}$ in \eqref{H1st}. In particular, we are wondering whether the mathematical techniques used in Ref.~\cite{Seiringer11} to estimate the accuracy of Bogoliubov's theory rigorously can be adapted to the context of RDMFT. In that respect, it is worth noticing that our result \eqref{Fmin} with phases $\sigma_{\bd{p}} =\mbox{sgn}(W_{\bd{p}})$ (or any other choice of phases) represents a universal upper bound to the functional of the full pair interaction $\hat{W}$ in the entire domain. This is due to the following two reasons. First, \eqref{Fmin} is found through a variational ansatz and second this ansatz of paring states has only an overlap with the pairing interaction Hamiltonian $\hat{W}_P$ (recall Eq.~\eqref{WP}). Proving in a \emph{mathematically rigorous} way the BEC force might be particularly challenging. Even if a lower and upper bound can be proven for the difference between the Bogoliubov and the full functional $\F_{\hat{W}}$, the gradient cannot easily be controlled since $\F_{\hat{W}}$ may (at least in principle) strongly oscillated between both bounds. A more promising approach would be to exploit the one-to-one correspondence established through the constrained search formalism between $\bd{n}$ and the Bogoliubov trial state, $\bd{n} \leftrightarrow \hat{U}_{\bd{n}} \ket{N}$. The idea for obtaining higher order terms of \eqref{FBog} would be to conjugate in a first step $\hat{W}$, $\hat{U}_{\bd{n}} \hat{W}  \hat{U}_{\bd{n}}^\dagger$, and afterwards apply conventional perturbation theory with respect to the resulting small terms in $\hat{U}_{\bd{n}} \hat{W}  \hat{U}_{\bd{n}}^\dagger$ (which are dropped within Bogoliubov theory). These higher order corrections would then allow one to prove the BEC force in a mathematically comprehensive way by generalizing techniques that have been developed in \cite{Benavides20} in the context of the Hubbard dimer.

\begin{acknowledgements}
C.S. acknowledges financial support from the Deutsche Forschungsgemeinschaft (Grant SCHI 1476/1-1).
\end{acknowledgements}

\appendix

\section{Functional and ground state energy for the dilute Bose gas}\label{app:Fdil}
In this section we derive the functional $\mathcal{F}(\thickbar{\bd{n}})$ for the dilute Bose gas in 3D and use this result to obtain the well-known expression for the ground state energy.

Let us first emphasize that replacing already in Eq.~\eqref{FBog} all Fourier coefficients $W_{\bd{p}}$ by $W_{\mathbf{0}}$
would make the respective sum divergent. Instead, we rewrite \eqref{FBog} as
\begin{eqnarray}
\mathcal{F}(\thickbar{\bd{n}}) &=& n\sum_{\bd{p}\neq 0}W_{\bd{p}}\left(\thickbar{n}_{\bd{p}}-\sqrt{\thickbar{n}_{\bd{p}}(\thickbar{n}_{\bd{p}}+1)} + \frac{nW_{\bd{p}}m}{p^2} \right) \nonumber \\
&& \quad - \sum_{\bd{p}\neq 0}\frac{n^2W_{\bd{p}}^2m}{p^2}
\end{eqnarray}
since then one is allowed to replace $W_{\bd{p}}$ by $W_{\mathbf{0}}$ in the first term. This yields (also replacing the sum by an integral)
\begin{eqnarray}\label{Fphysnp}
\mathcal{F}(\thickbar{\bd{n}}) &=& \frac{V}{2\pi^2}n\int_0^\infty\rmd p\,p^2 W_{\mathbf{0}} \left(\thickbar{n}_{\bd{p}}-\sqrt{\thickbar{n}_{\bd{p}}(\thickbar{n}_{\bd{p}}+1)}\right. \nonumber \\
&& \quad\left.+ \frac{nW_{\mathbf{0}}m}{p^2} \right)- \sum_{\bd{p}\neq 0}\frac{n^2W_{\bd{p}}^2m}{p^2}\\
&=& \frac{V}{2\pi^2}(2m)^{3/2}\frac{2}{3}\sqrt{2}(nW_{\mathbf{0}})^{5/2} - \sum_{\bd{p}\neq 0}\frac{n^2W_{\bd{p}}^2m}{p^2}\,. \nonumber
\end{eqnarray}
The second term can be rewritten in terms of $a_1$ given by Eq.~\eqref{a0}. Including also the constant term which we neglected so far and replacing $W_{\mathbf{0}}$ by $a_0$ through Eq.~\eqref{a0} yields
\begin{eqnarray}
\mathcal{F}(\thickbar{\bd{n}}) &=& \frac{nN2\pi}{m}a_0 + \frac{nN\sqrt{\pi}}{m}\frac{128}{3}a_0(na_0^3)^{1/2} + \frac{4\pi nN}{m}a_1\nonumber \\
&=& \frac{2\pi nN}{m}\left(a_0 + \frac{64}{3\sqrt{\pi}}a_0(na_0^3)^{1/2} +2a_1\right)\,.
\end{eqnarray}

As a consistency test we start now from Eq.~\eqref{Fphysnp} and add the kinetic energy.
The second term in Eq.~\eqref{Fphysnp} can be split into two parts such that it cancels the divergence in the integral for the kinetic energy as follows:
\begin{eqnarray}
E &=& \sum_{\bd{p}\neq 0}\frac{p^2}{2m}n_{\bd{p}} + \mathcal{F}(\thickbar{\bd{n}})\\
&=& \frac{nNW_{\mathbf{0}}}{2} + \frac{V}{2\pi^2}(2m)^{3/2}\frac{2}{3}\sqrt{2}(nW_{\mathbf{0}})^{5/2} \nonumber \\
&&\quad+ \frac{V}{2\pi^2}\int_0^\infty\rmd p\,p^2\left(\frac{p^2}{2m}n_{\bd{p}} - \frac{n^2W_{\mathbf{0}}^2m}{2p^2}\right) - \sum_{\bd{p}\neq 0}\frac{n^2W_{\bd{p}}^2m}{2p^2}\nonumber \\
&=& \frac{nNW_{\mathbf{0}}}{2} +\frac{V}{2\pi^2}(2m)^{3/2}\frac{4}{15}\sqrt{2}(nW_{\mathbf{0}})^{5/2} - \sum_{\bd{p}\neq 0}\frac{n^2W_{\bd{p}}^2m}{2p^2}\,. \nonumber
\end{eqnarray}
Inserting $a_0$ and $a_1$ leads to the ground state energy
\begin{equation}
E = \frac{4\pi Nn}{2m}(a_0+a_1) + \frac{4\pi Nn}{2m}a_0\frac{128}{15\sqrt{\pi}}(na_0^3)^{1/2}
\end{equation}
which is in agreement with Ref.~\cite{Brueckner57}.

\section{BEC force for the dilute Bose gas}\label{app:forcedil}
We calculate the derivative of $\mathcal{F}$ with respect to the distance along a straight path denoted by $s$ towards complete BEC starting at the occupation number vector $\thickbar{\bd{n}}$. Then, $\thickbar{n}_{\bd{p}}(t) = \thickbar{n}_{\bd{p}}(1-t)$ and for $t\approx 1$ or equivalently $D(t)\ll 1$ we can approximate
\begin{eqnarray}\label{dFsum}
\left.\frac{\rmd \mathcal{F}(\thickbar{\bd{n}})}{\rmd D}\right\vert_s &=& \frac{1}{D(t)} \sum_{\bd{p}\neq 0}nW_{\bd{p}}\thickbar{n}_{\bd{p}}(t)\left(1 - \frac{2\thickbar{n}_{\bd{p}}(t)+1}{2\sqrt{\thickbar{n}_{\bd{p}}(t)(\thickbar{n}_{\bd{p}}(t)+1)}}\right)\nonumber \\
&\approx& -\left(\frac{n}{2\sqrt{D(0)}}\sum_{\bd{p}\neq 0}W_{\bd{p}}\sqrt{\thickbar{n}_{\bd{p}}}\right)\frac{1}{\sqrt{D(t)}}\,.
\end{eqnarray}
The summation in Eq.~\eqref{dFsum} can be replaced by an integral ($\sum_{\bd{p}}\to \frac{V}{(2\pi)^3}\int \rmd^3\bd{p}$) in the thermodynamic limit where $N\to \infty$, $V\to\infty$ and $n=N/V=\mathrm{cst.}$. To evaluate the integral over the momentum $\bd{p}$ we rewrite Eq.~\eqref{dFsum} as follows:
\begin{equation}\label{F_nbar_a1}
\begin{split}
\left.\frac{\rmd \mathcal{F}(\thickbar{\bd{n}})}{\rmd D}\right\vert_s &\approx -\frac{n}{4\pi^2\sqrt{D(0)}}\int_0^\infty\rmd p\,p^2 \left(W_{\bd{p}}\sqrt{\thickbar{n}_{\bd{p}}} \right.\\\
&\quad\left.- \frac{(nW_{\bd{p}})^2m}{p^2}\right)\frac{1}{\sqrt{D(t)}}\\\
&\quad- \frac{1}{2\sqrt{D(0)}}\sum_{\bd{p}\neq 0}\frac{(nW_{\bd{p}})^2m}{p^2}\frac{1}{\sqrt{D(t)}}
\end{split}
\end{equation}
such that the integral over $p$ is converging after replacing $W_{\bd{p}}$ by the constant value $W_{\mathbf{0}}$. The first two terms in the Born series for the scattering length $a$ for identical particles are given by Eq.~\eqref{a0}. Thus, the summation in the second line of Eq.~\eqref{F_nbar_a1} can be identified with $a_1$ and the result of the integration in the first line will depend on $W_{\mathbf{0}}$ which can be replaced by $a_0$ through Eq.~\eqref{a0}. Since the integral can only be evaluated numerically we define a positive constant $\eta(a_0, n, m)$ for its value and obtain for the derivative of $\mathcal{F}$ along the path $s$:
\begin{equation}\label{F_nbar_result}
\left.\frac{\rmd \mathcal{F}(\thickbar{\bd{n}})}{\rmd D}\right\vert_s\approx \frac{N\eta(a_0, n, m)}{\sqrt{D(t)}} + \frac{2\pi nNa_1}{m\sqrt{D(0)}}\frac{1}{\sqrt{D(t)}}\,.
\end{equation}

\section{Universal functional for the Bose-Hubbard model}\label{app:signs}
In this section we solve the minimization in Eq.~\eqref{Fmin} for any pair of occupation numbers $(n_1, n_2)$ for the Bose-Hubbard model with $N$ bosons on $L=5$ lattice sites and $U>0$. Since in that case the Fourier coefficients $W_p= \mathrm{sgn}(U)$ are independent of the momentum $p$, they can be pulled out of the summation over $p$. The four different combinations of the signs are $(\sigma_1, \sigma_2)=(+,+), (+, -), (-,+), (-, -)$ and the four corresponding functionals are denoted by $\mathcal{F}_{(\sigma_1, \sigma_2)}$. The functional $\mathcal{F}_{(-,-)}$  can be neglected in the following discussion since it comprises only positive terms and thus $\mathcal{F}_{(\sigma_1,\sigma_2)}  \leq \mathcal{F}_{(-,-)}$ for all $(\sigma_1,\sigma_2)$. The remaining three functionals $\mathcal{F}_{\sigma_1, \sigma_2}$ are then split into $\mathcal{F}_{(\sigma_1, \sigma_2)} = 2(\mathcal{F}^{(1)} + \mathcal{F}_{(\sigma_1, \sigma_2)}^{(2)})/L$ where $\mathcal{F}^{(1)}$ is independent of the choice of signs $(\sigma_1, \sigma_2)$. Therefore, to find the minimizing configuration for any $\bd{n} \in \triangle$ we only have to compare
\begin{equation}
\begin{split}
\mathcal{F}_{(+,+)}^{(2)} &= -\sum_{\nu=1}^2\left(n_{\mathbf{0}} - \sum_{\nu=1}^2\sqrt{n_\nu(n_\nu+1)}\right)\sqrt{n_\nu(n_\nu+1)}\\\
\mathcal{F}_{(+,-)}^{(2)} &= -\left(n_{\mathbf{0}}- \sqrt{n_1(n_1+1)}+\sqrt{n_2(n_2+1)}\right)\\\
&\quad\times\left(\sqrt{n_1(n_1+1)} - \sqrt{n_2(n_2+1)}\right)
\end{split}
\end{equation}
and the third functional $\mathcal{F}^{(2)}_{(-,+)}$ follows from $\mathcal{F}^{(2)}_{(-,+)}$ by replacing everywhere $1\leftrightarrow 2$.
The minimizing configuration $(\sigma_1, \sigma_2)$  can then easily be determined analytically leading to the cells shown in Fig.~\ref{fig:S}.
There, the black point in the middle marks the distinctive occupation number vector for which all three functionals take the same value  $\mathcal{F}_{(+,+)}=\mathcal{F}_{(-,+)}=\mathcal{F}_{(+,-)}$. It is given by
\begin{equation}
\tilde{n}\equiv n_1=n_2=\frac{1}{6}\left(1+2N-\sqrt{1+N(4+N)}\right) \,.
\end{equation}
The border between regions $(-,+)$ and $(+,-)$ is determined by
\begin{equation}
n_2=n_1 \geq \tilde{n}\,.
\end{equation}
The border separating region $(+,+)$ and $(+,-)$ is obtained from $\mathcal{F}_{(+,+)}(\bd{n})=\mathcal{F}_{(+,-)}(\bd{n})$, leading to
\begin{equation}
n_2 = \frac{1}{2}\left(N - 2\left(n_1+\sqrt{n_1(n_1+1)}\right)\right)\,,\quad n_1\geq \tilde{n}\,.
\end{equation}
The solution for $\mathcal{F}_{(+,+)}=\mathcal{F}_{(-,+)}$ is obtained by exchanging the two occupation numbers $n_1$ and $n_2$ in the result for $\mathcal{F}_{(+,+)}=\mathcal{F}_{(+,-)}$.

\bibliography{Refs}

\end{document}